\documentclass[preprint,showpacs,showkeys,aps,prd,amsfonts,eqsecnum,nofootinbib]{revtex4}

\usepackage{graphicx,epsfig}
\usepackage[dvips]{color}

\begin{document}
\preprint{CUMQ/HEP 140\\,~~\\IZTECH-P/2006--04}
%
%
\title{\Large NEUTRALINO DARK MATTER IN THE LEFT-RIGHT SUPERSYMMETRIC MODEL}

\author{Durmu{\c s} A. Demir$^{a}$}\email[]{demir@physics.iztech.edu.tr}
\author{Mariana Frank$^{b}$}\email[]{mfrank@vax2.concordia.ca}
\author{Ismail Turan$^{b}$}\email[]{ituran@physics.concordia.ca}

\affiliation{$^{a}$ Department of Physics, Izmir Institute of
Technology, IZTECH, TR35430, Izmir, Turkey}

\affiliation{$^{b}$
Department of Physics, Concordia University, 7141 Sherbrooke
Street West, Montreal, Quebec, CANADA H4B 1R6}
\date{\today}

\begin{abstract}

We study the neutralino sector of the left-right supersymmetric
model. In addition to the possibilities available in the minimal
supersymmetric model, the  neutralino states can be superpartners
of the $U(1)_{B-L}$ gauge boson, the neutral $SU(2)_R$ neutral
gauge boson, or of the Higgs triplets. We analyze neutralino
masses and determine the parameter regions for which the lightest
neutralino can be one of the new pure states. We then calculate
the relic density of the dark matter for each of these states and 
impose the constraints coming from the $\rho$ parameter, 
the anomalous magnetic moment of the muon,  
$ b\to s \gamma$, as well as general supersymmetric mass bounds. The
lightest neutralino can be the bino, or the right-wino, or the
neutral triplet higgsino, all of  which have different couplings to the
standard model particles from the usual neutralinos. A light bino satisfies all
the experimental constraints and would be the preferred dark matter candidate for 
light supersymmetric scalar  masses, while the right-wino would be favored by 
intermediate supersymmetric mass scales. The neutral triplet Higgs fermion 
satisfies the experimental bounds only in a small region of the parameter space, 
for intermediate to heavy supersymmetric scalar masses.
\pacs{12.60.Cr, 12.60.Fr, 14.80.Ly}
\keywords{neutralinos, dark matter, left right supersymmetry}
\end{abstract}
\maketitle
\section{Introduction}\label{sec:intro}

Recent measurements from the Wilkinson Microwave Anisotropy Probe
(WMAP) satellite \cite{Peiris:2003ff} have shown that the cold
dark matter (CDM) abundance in the universe is $\Omega_{CDM}
h^2=0.1126^{+0.00805}_{-0.00904}$, where $\Omega_{CDM}$ is the
density of the CDM species versus the critical density, and $h$ is
the present value of the Hubble parameter. This corresponds to
about $22\%$ of the energy density of the universe being present
in the form of dark matter.

The question of its nature and composition has been open problem
for some time. A candidate for dark matter must be stable, or
long-living; it must be electrically and color neutral, as
required by astrophysical constraints; and it must be
non-relativistic, which means that it should be massive. Since the
relic density can be connected to the  (thermally averaged)
annihilation cross section of a dark matter candidate, its value
indicates that these particles are interacting weakly. That is,
dark matter candidates must be weakly interacting massive
particles. Scenarios beyond the standard model provide several
such exotic particles.  Supersymmetry, in particular,  provides
the lightest neutralino as the lightest supersymmetric particle
(LSP), which must be stable if R-parity is conserved.\footnote{
Gravitinos and Axinos have also been studies as dark matter
candidates, but we shall not consider this possibility here.}

The MSSM Lagrangian, based on the symmetry group  $SU(3)_C \otimes
SU(2)_L \otimes U(1)_Y$ is invariant under the R-parity discrete
symmetry, under which all standard model particles are even, and
their superpartners are odd. This parity is an essential element
of the  MSSM, and forbids baryon and lepton number violating
renormalizable couplings in the superpotential, which would lead
to fast proton decay. Thus the R-parity is natural, and so is the
possibility of having an LSP which emerges as a natural candidate
for dark matter. In most supersymmetric scenarios, the LSP is the
lightest neutralino \cite{lsp}.

Unfortunately there are  problems with this CDM assignment  in
MSSM \cite{history}. If one calculates the relic abundance of the
lightest neutralino assumed to be a pure bino (the superpartner of
the gauge boson of the $U(1)_Y$ gauge group), the relic density is
too large (see  \cite{Arkani-Hamed:2006mb} and references therein
for a detailed discussion). To avoid this problem, one considers
that there are several other supersymmetric particles of masses
close to the LSP. If such sparticles exist, their co-annihilation
with the LSP leads to a reduction of the LSP relic abundance
\cite{Griest:1990kh}. Extensive studies of such phenomena exist,
and several publicly available codes such as \texttt {DarkSUSY}
\cite{Gondolo:2004sc} and \texttt{micrOMEGA}
\cite{Belanger:2004yn} , which include all the relevant
coannihilation channels,  are available for calculations within
the MSSM. Cosmological and phenomenological constraints (such as
$b\to s \gamma$ \cite{Gomez:2000sj}, $(g-2)_{\mu}$
\cite{Chattopadhyay:2003xi}, and Higgs mass bounds
\cite{Barger:1998ga}) are also included. We could of course also
abandon the pure bino state, and thus the constrained MSSM (CMSSM)
which predicts it, and explore other versions of the lightest
neutralino. However, this does not solve the problem because if
one allows the lightest neutralino to be mostly left-wino, or
higgsino, the relic density becomes too small. Ways to avoid this
exist as well, such as considering non-universal gaugino masses, 
abandoning the pure states and examining mixed states (such as bino-higgsino), 
or more annihilations \cite{Baer:2005jq}.

An alternative option is to explore scenarios beyond the MSSM. The
extensions of the MSSM, motivated by the need to a dynamical
solution to the naturalness problem associated with the $\mu$
parameter, offer novel CDM candidates not necessarily belonging to
gaugino or Higgsino sectors of the MSSM. In next-to-minimal model
(NMSSM)  \cite{Belanger:2005kh}, for instance, the neutralino can
be extremely light because it is dominated by its singlino
component, which does not couple to SM particles (except for Higgs
doublets). The existence of a very light CP-odd Higgs boson
provides the possibility for this neutralino to annihilate
sufficiently to avoid being overproduced in the early universe.
Similar features are also found in U(1)$^{\prime}$ models
\cite{deCarlos:1997yv} or in more general extensions
\cite{Barger:2005hb}. This raises the hope that models beyond
MSSM, preferably comprising the neutrino masses as well, can
provide viable scenarios for alternative neutralinos which have
different features than those in MSSM.

We propose here to look at the left-right supersymmetric model
(LRSUSY) for candidates for dark matter. The advantages of the
LRSUSY are many, such as combining supersymmetry with right handed symmetry, and
thus providing a mechanism for  neutrino masses with the seesaw
within a supersymmetric scenario. From the point of view of dark matter, 
the LRSUSY model is interesting because, unlike in MSSM,  the usual explicit R-parity violating terms are forbidden in the superpotential by the symmetries of the model \cite{Aulakh:1999pz}.\footnote{Note however that the R-parity can be broken spontaneously \cite{Huitu:1994zm}.}  Thus the model naturally predicts a lightest supersymmetrical particle. 

The LRSUSY model emerges
naturally in some superstring theories  \cite{Frank:2005rb}, or in
the breaking of such SUSY GUT scenarios, such as $SO(10)$
\cite{Dutta:2005ni}. LRSUSY is based on the symmetry group
$SU(3)\otimes SU(2)_L\otimes SU(2)_R\otimes U(1)_{V=B-L}$ and thus
provides several new sources for exotic neutralinos, in the
(neutral) partners of the $SU(2)_R$ or $U(1)_{B-L}$ gauge bosons,
as well as the fermionic partners of several Higgs bosons. We
analyze the candidates for dark matter in this model to see if
they can avoid the problems that plague MSSM. After briefly
reviewing the model in Section~\ref{model}, we perform a comprehensive
study of neutralino mass eigenstates in Section~\ref{classification}, followed by a
calculation of relic density in Section~\ref{relicdensity}. We look at the
simplest scenario, that of pure states and avoid co-annihilation
with other supersymmetric particles by choosing the supersymmetric
masses accordingly. We include constraints from the WMAP, $b
\rightarrow s \gamma$, $\Delta\rho$, and muon $g-2$ as well as from experimental
bounds on supersymmetric masses from direct collider searches.


\section{The LRSUSY Model}
\label{model}

\noindent The most general superpotential for the group $SU(3)\otimes
SU(2)_L\otimes SU(2)_R\otimes U(1)_{V=B-L}$ is \cite{Francis:1990pi}  :
\begin{eqnarray}
\label{superpotential}
W & = & {\bf Y}_{Q}^{(i)} Q^T\Phi_{i}i \tau_{2}Q^{c} + {\bf Y}_{L}^{(i)}
L^T \Phi_{i}i \tau_{2}L^{c} + i({\bf Y}_{LR}L^T\tau_{2} \delta_L L +
{\bf Y}_{LR}L^{cT}\tau_{2}
\Delta_R L^{c}) \nonumber \\
& & + \mu_{LR}\left [{\rm Tr} (\Delta_L  \delta_L +\Delta_R
\delta_R)\right] + \mu_{ij}{\rm Tr}(i\tau_{2}\Phi^{T}_{i} i\tau_{2} \Phi_{j})
+W_{NR},
\end{eqnarray}
where $W_{NR}$ denotes (possible) non-renormalizable terms arising
from higher scale physics or Planck scale effects \cite{Aulakh:1998nn, Chacko:1997cm}. Here the matter fields
are defined as
\begin{eqnarray}
Q&=&\left (\begin{array}{c}
u\\ d \end{array} \right ) \sim \left ( 3,2, 1, 1/3 \right ),~~
Q^c=\left (\begin{array}{c}
d^c\\u^c \end{array} \right ) \sim \left ( 3^{\ast},1, 2, -1/3
\right ),\nonumber \\
L&=&\left (\begin{array}{c}
\nu\\ e\end{array}\right ) \sim\left( 1,2, 1, -1 \right),~~
L^c=\left (\begin{array}{c}
e^c \\ \nu^c \end{array}\right ) \sim \left ( 1,1, 2, 1 \right ),
\end{eqnarray}
where the numbers in the brackets denote the quantum numbers under
$SU(3)_C \otimes SU(2)_L \otimes SU(2)_R \otimes U(1)_{B-L}$. The Higgs
sector consists of the bidoublet and triplet Higgs superfields:
\begin{eqnarray}
\displaystyle
\Phi_1 &&\hspace*{-0.3cm}= \left (\begin{array}{cc}
\phi^0_{11}&\phi^+_{11}\\ \phi_{12}^-& \phi_{12}^0
\end{array}\right)\hspace*{0.5cm} \sim \left (1,2,2,0 \right),\hspace*{0.62cm}
\Phi_2=\left (\begin{array}{cc}
\phi^0_{21}&\phi^+_{21}\\ \phi_{22}^-& \phi_{22}^0
\end{array}\right)\hspace*{0.1cm} \sim \left (1,2,2,0 \right), \nonumber \\
\Delta_{L} &&\hspace*{-0.3cm} = \left(\begin{array}{cc}
\frac {\Delta_L^-}{\sqrt{2}}&\Delta_L^0\\
\Delta_{L}^{--}&-\frac{\Delta_L^-}{\sqrt{2}}
\end{array}\right) \sim (1,3,1,-2),~~~\delta_{L}  =
\left(\begin{array}{cc}
\frac {\delta_L^+}{\sqrt{2}}&\delta_L^{++}\\
\delta_{L}^{0}&-\frac{\delta_L^+}{\sqrt{2}}
\end{array}\right) \sim (1,3,1,2),\nonumber \\
\Delta_{R} &&\hspace*{-0.3cm} =
\left(\begin{array}{cc}
\frac {\Delta_R^-}{\sqrt{2}}&\Delta_R^0\\
\Delta_{R}^{--}&-\frac{\Delta_R^-}{\sqrt{2}}
\end{array}\right) \sim (1,1,3,-2),~~~\delta_{R}  =
\left(\begin{array}{cc}
\frac {\delta_R^+}{\sqrt{2}}&\delta_R^{++}\\
\delta_{R}^{0}&-\frac{\delta_R^+}{\sqrt{2}}
\end{array}\right) \sim (1,1,3,2).
\end{eqnarray}
where $\Delta_L$ and $\delta_R$ are introduced in the model to cancel
the fermionic anomalies introduced by the fermionic partners of $\delta_L$ and $\Delta_R$.
The vev's of the Higgs fields in the LRSUSY can be chosen
\begin{eqnarray}
\langle \Phi_1 \rangle&&\hspace*{-0.3cm} = \left (\begin{array}{cc}
\kappa_1&0\\0& 0
\end{array}\right),~~~\langle \Phi_2 \rangle = \left (\begin{array}{cc}
0 &0\\0&\kappa_2
\end{array}\right), ~~~\langle \Delta_{L} \rangle = \left(\begin{array}{cc}
0&v_{\Delta_L}\\0&0
\end{array}\right), \nonumber \\
\langle ~\delta_{L} \rangle&&\hspace*{-0.3cm} = \left
(\begin{array}{cc} 0&0\\v_{\delta_L}&0
\end{array}\right),\hspace*{0.4cm}~\!\!\!\!\!\langle \Delta_{R} \rangle = \left
(\begin{array}{cc} 0&v_{\Delta_R}\\0&0
\end{array}\right),\hspace*{0.5cm} ~\!\!\!\!\!\langle \delta_{R} \rangle = \left
(\begin{array}{cc} 0&0\\v_{\delta_R}&0
\end{array}\right).
\end{eqnarray}
Some comments and explanations about the vev's chosen are
required: $\kappa_1$ and $\kappa_2$ are the vev's of the MSSM-like
Higgs bosons. They are responsible for giving masses to the quarks
and leptons and they also contribute to $M_{W_L}$.  We take the vev's
of $\phi_{12}^0$ and $\phi_{21}^0$ to be zero because they induce
FCNC at tree level (in both the leptonic and hadronic systems), as
well as being responsible for $W_L -W_R$ mixing. The vev's could
also have a phase which induces CP violation, which is severely
restricted in the kaon system. The non-MSSM Higgs vev's,
$v_{\delta_L}$ and $ v_{\Delta_R}$ are responsible for neutrino
masses.  For one generation (see  \cite{Mohapatra:1980yp}, also 
 \cite{Aulakh:1997ba,Aulakh:1998nn,Frank:2002hk})
\begin{equation}
m_{\nu}=Y_{LR}v_{\delta_L}-\frac{(Y_L^{(\nu)})^2 \kappa_1^2}{Y_{LR}\, v_{\Delta_R}}
\end{equation}
where $v_{\delta_L}$ must be very  small and $v_{\Delta_R}$ large,
phenomenologically . In addition, $v_{\Delta_L}$ and $v_{\delta_L}$
enter in the formula for the mass of $W_L$ (or the $\rho$
parameter), while $v_{\Delta_R}, v_{\delta_R}$ enter in the
formula for the mass of $W_R$. It is thus justified to take
$v_{\Delta_L}, v_{\delta_L}$ to be negligibly small. For
$v_{\Delta_R} $ there are two possibilities: either  $v_{\Delta_R}
$ is $\approx 10^{13}$ GeV \cite{Frank:2002hk, Aulakh:1997fq},
which supports the seesaw mechanism, leptogenesis and provides
masses for the light neutrinos in agreement with experimental
constraints, but offers no hope to see right-handed particles; or
$v_{\Delta_R} $ is $\approx 1- 10$ TeV, but one must introduce
something else (generally an intermediate scale, or an extra
symmetry) to make the neutrinos light  \cite{Aulakh:1997fq,Aulakh:1998nn,Aulakh:1999pz}.
 Note that both $v_{\Delta_R}$ and $v_{\delta_R}$ contribute to the mass
of $W_R$ \cite{Frank:2003un}, but only one needs to be heavy.
Since $v_{\Delta_R}$ is responsible for heavy right-handed
neutrino masses, it must be large. Thus $v_{\delta_R}$  is not
constrained by the data. The mass term for neutralinos is given by
\begin{eqnarray}
{\cal L}_N=-\frac{1}{2} {\psi^0}^T Y \psi^0  + {\rm H.c.} \ ,
\end{eqnarray}
where $\psi^0=(-i \lambda_V, -i \lambda_L^0,
\tilde{\phi}_{11}^0,\tilde{\phi}_{22}^0,
\tilde{\Delta}_R^0,\tilde{\delta}_R^0,-i \lambda_R^0,
\tilde{\phi}_{12}^0, \tilde{\phi}^0_{21} )^T $. Here $\lambda_V$
is the $U(1)_{B-L}$ bino, $\lambda_L^0$ the
  left-handed neutral ($SU(2)_L$) wino, and $\lambda_R^0$
the right-handed neutral ($SU(2)_R$) wino. The rest of the fields
are higgsinos.  The neutralino mass matrix $Y$ is equal 
to \cite{Frank:2005vd}
\begin{eqnarray}
\label{Ymatrix}
\!\!\!\!\!\!\!\!\!\!\!\!\!\!Y=\left( \begin{array}{c|cccc|ccccc}
  & -i \lambda_V & -i \lambda_L^0 & \tilde{\phi}_{11}^0 & \tilde{\phi}_{22}^0 &
\tilde{\Delta}_R^0 & \tilde{\delta}_R^0 & -i \lambda_R^0 &
\tilde{\phi}_{12}^0 & \tilde{\phi}^0_{21}\\
\hline
-i \lambda_V & M_V & 0 & 0 & 0 &  -\sqrt{2} g_V v_{\Delta} &
\sqrt{2} g_V v_{\delta} &0 & 0  & 0 \\
-i \lambda_L^0 & 0 & M_L  & \frac{g_L \kappa_1}{\sqrt{2}} &
-\frac{g_L \kappa_2}{\sqrt{2}} & 0 & 0 & 0 & 0& 0\\
\tilde{\phi}_{11}^0 & 0 & \frac{g_L \kappa_1}{\sqrt{2}} & 0 & -\mu_1
& 0 & 0 & -\frac{g_R \kappa_1}{\sqrt{2}} & 0 &0\\
  \tilde{\phi}_{22}^0 & 0 & -\frac{g_L \kappa_2}{\sqrt{2}} & -\mu_1& 0
& 0 & 0 & \frac{g_R \kappa_2}{\sqrt{2}} & 0 & 0  \\
\hline
  \tilde{\Delta}_R^0 & -\sqrt{2} g_V v_{\Delta} & 0 & 0 & 0 & 0 &
\mu_2 & \sqrt{2} g_R v_{\Delta} & 0 & 0 \\
  \tilde{\delta}_R^0 &  \sqrt{2} g_V v_{\delta} & 0 & 0 & 0 & \mu_2 &
0 & -\sqrt{2} g_R v_{\delta} & 0 & 0\\
  -i \lambda_R^0 &  0 & 0 & -\frac{g_R \kappa_1}{\sqrt{2}} & \frac{g_R
\kappa_2}{\sqrt{2}} & \sqrt{2} g_R v_{\Delta} & -\sqrt{2} g_R
v_{\delta} & M_R & 0 &0    \\
\tilde{\phi}_{12}^0 & 0 & 0 & 0 & 0 & 0 & 0 & 0 & 0 & -\mu_1 \\
\tilde{\phi}^0_{21} & 0 & 0 & 0 & 0 & 0 & 0 & 0 & -\mu_1 & 0
                \end{array}
          \right )
\end{eqnarray}
where $\mu_{ij}=\mu_1,\,\,(i\ne j)$ and $\mu_{LR}\equiv \mu_2$ are
assumed.\footnote{From now on we drop the subscript ``$R$" from
$v_{\Delta _R},~v_{\delta_R}$.} The upper $4\times 4$ part of the matrix contains
MSSM-like states. There are still too many unknown parameters and
for further simplification one can define the followings:
\begin{eqnarray}
&&g_L = g_R \equiv g\,,\;\;\;g_V =\frac{e}{\sqrt{\cos 2\theta_W}}\,,\;\;\sqrt{\kappa_1^2+\kappa^2_2}=174\, {\rm GeV}\,,\,\, \tan\beta =
\frac{\kappa_2}{\kappa_1}.
\end{eqnarray}
If we assume that GUT relations between gaugino masses hold, we
can simplify the parameters further, but for now we  keep them
general and discuss specific scenarios later.

Two roots of the characteristic
equation for $Y$ are already known, $\pm\mu_1$, and we are left with
a $7^{th}$ order equation
to solve.  There is no exact solution and a numerical
or semi-exact approach is necessary. Whereas there are constraints on
$M_{W_R}$ and $M_{Z_R}$ \cite{Eidelman:2004wy}, there are no constraints on $M_V$, the
$U(1)_{B-L}$ gaugino mass, or on $M_{L(R)}$, the $SU(2)_{L(R)}$ gaugino mass parameter.

Basically the number of free parameters can be chosen as
$M_V,\,M_L,\,M_R,\,v_\Delta,\,v_\delta, \,\mu_1, \, \mu_2$, and
$\tan\beta$. So, we could look at under what circumstances the
lightest neutralino becomes mainly pure state (bino, left-wino,
right-wino, or higgsino) or a mixed state as bino--higgsino and
right-wino--higgsino.

\section{The Classification of Scenarios
\label{classification}} Starting from the neutralino mass matrix,
we can classify the following cases:\footnote{We set
$v_{\Delta}=1.5$ TeV for all the cases unless otherwise stated.
Note also that there will be no rigorous model constraint analysis
on the masses of the LSP in this section. We leave that to the
next Section.}

\begin{enumerate}

\item { The lightest neutralino is mostly
bino ($\lambda_V$ in our notation). Note that this the the $B-L$
gaugino, not to be confused with the $U(1)_Y$ MSSM bino. To obtain
this bino as the lightest state, we must have $M_V \approx 50-100$
GeV and $v_{\delta} \approx 50-100$ GeV (both light) while $\mu_2
\approx 1$ TeV or larger (heavy). Decreasing $\mu_2$ to $200-500$ GeV, the
lightest state becomes a mixture of $\lambda_V$ and
$\tilde
\delta_R^0$.}

\item { The lightest neutralino is mostly right-handed
wino ($\lambda_R^0$ in our notation).  To obtain this, the masses of
the other two gauginos must be larger than that of $\lambda_R^0$:
$M_V, \,M_L \approx 600$ TeV and $v_{\delta} \approx 50-100$ GeV is
light; while $\mu_1 \approx 1$ TeV, $\mu_2 \approx 3-5$ TeV.
Decreasing $\mu_2$ to $1$ TeV, the lightest state becomes a mixture
of $\lambda_R^0$, $\lambda_V$ and
$\tilde \delta_R^0$.}

\item{ The
lightest neutralino is the right-handed higgsino $\tilde
\delta_R^0$. This scenario is obtained for a large range of
parameters, as long as both $\mu_2$ and $v_{\delta}$ are small,
$\approx 200$ GeV or larger, and $(0 , 100)$ GeV, respectively. In
fact, this requirement is satisfied for a wide portion of the
parameter space, as long as $\mu_2$ and $v_{\delta}$ are smaller
than other parameters in the theory. This scenario is interesting
since $\tilde \delta_R^0$ does not couple to any Standard Model
particles (or their SUSY partners). As explained previously,
$v_{\delta}$ is not constrained by the data. }

\item{ The lightest neutralino
could be a mixture of $\lambda_V,\, \tilde \delta_R^0$ and $\tilde
\Delta_R^0$. In this case we can get eigenvectors and eigenvalues
analytically and calculate the relic density for the mixed state.
This is a generalization of the first case, where both
$\tilde \Delta_R^0$ and $\tilde \delta_R^0$ are included.
Here we can take advantageous ratios of the vevs
of $\tilde \delta_R^0$ and $\tilde \Delta_R^0$
 to get only one combination of $\tilde \delta_R^0$ and
$\tilde \Delta_R^0$ mixed with  the bino. This is the case
if we assume the vev's $v_{\delta}$ and $v_{\Delta}$ equal.}

\item{
The lightest neutralino could be a mixture of $\lambda_R^0$
(the right-handed wino), and the non-MSSM higgsinos
$\tilde \delta_R^0, \,\tilde \Delta_R^0$.  As
opposed to the $B-L$ bino, the right-handed wino mixes with MSSM
higgsinos as well, which will be considered in a separate scenario.
Again like in scenario (4), one can make
the right-wino couple with one combination of $\tilde \delta_R^0$
and $\tilde \Delta_R^0$ if their vev's are assumed to be equal.}

\item{
Finally, the lightest neutralino could be a mixture of
$\lambda_R^0$ with only the MSSM higgsinos, $\tilde \phi_{11}^0$
and $\tilde \phi_{22}^0$. This can be the case if both $M_R$ and
$\mu_1$ are small compared to $M_L,M_V,\mu_2$, and the vev's
$v_\Delta$ and $v_\delta$. One could consider $v_\Delta=v_\delta$
case and decouple $\tilde \delta_R^0$ and $\tilde \Delta_R^0$ by
taking $\mu_2$ to be large. }

\end{enumerate}

We are not interested in the MSSM lightest neutralino scenarios
(in which the left-wino or MSSM higgsino mixed with left wino are
the LSP), since these have already been studied. As far as we can
tell, scenarios 1-6 are the important (most striking)
possibilities.








\subsection{The lightest neutralino state}
\label{masses}
The following four sets of figures further illustrate
the scenarios mentioned above. We assume the composition of the
lightest state as (for the first three Scenarios)
\begin{equation}
\vert\tilde{\chi}^0_1\rangle=N_{11}\vert{\lambda}^0_L\rangle+N_{12}\vert{\lambda}^0_R\rangle+N_{13}\vert{\lambda}_V\rangle+N_{14}\vert\tilde
\phi^0_{11}\rangle+N_{15} \vert\tilde \phi^0_{22}\rangle
+N_{16}\vert\tilde{\Delta}^0_R\rangle+N_{17}\vert\tilde{\delta}^0_R\rangle.
\end{equation}

{\it First we look at the possibilities for Scenarios (1) and (3).
Here we want the bino to be the lightest, so we take $M_L$, $M_R$
and $\mu_1$ large. Accordingly, we need  $v_{\delta}$ to be small.
Varying $\mu_2$ will take us from a mostly bino lightest state to
a mostly $\tilde{\delta}_R^0$ lightest state.}

\begin{figure}[htb]
\vspace{-0.05in}
    \centerline{\hspace*{-0.2cm} \epsfxsize 3.57in
{\epsfbox{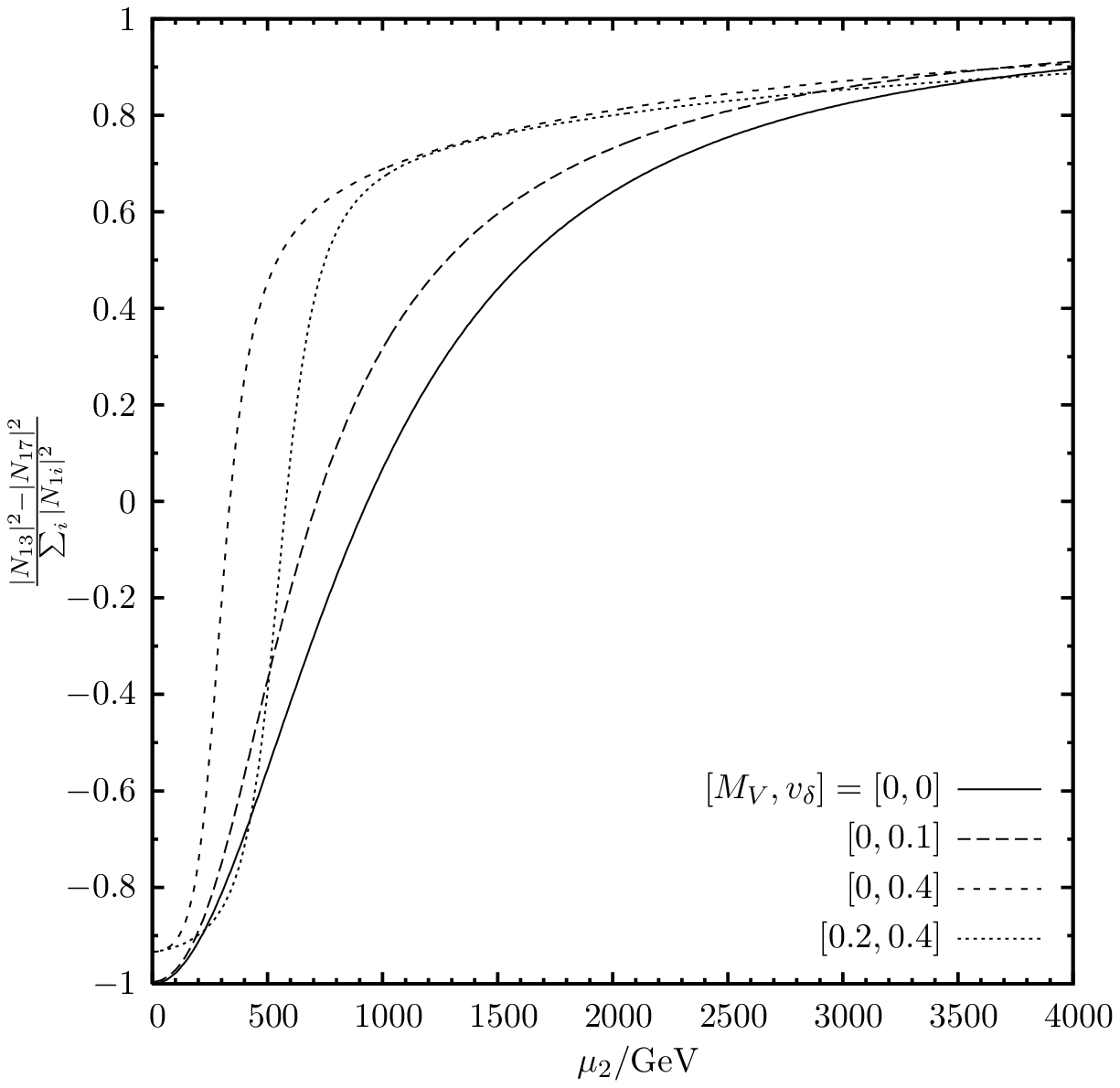}} \hspace{-0.65cm} \epsfxsize 3.5in
{\epsfbox{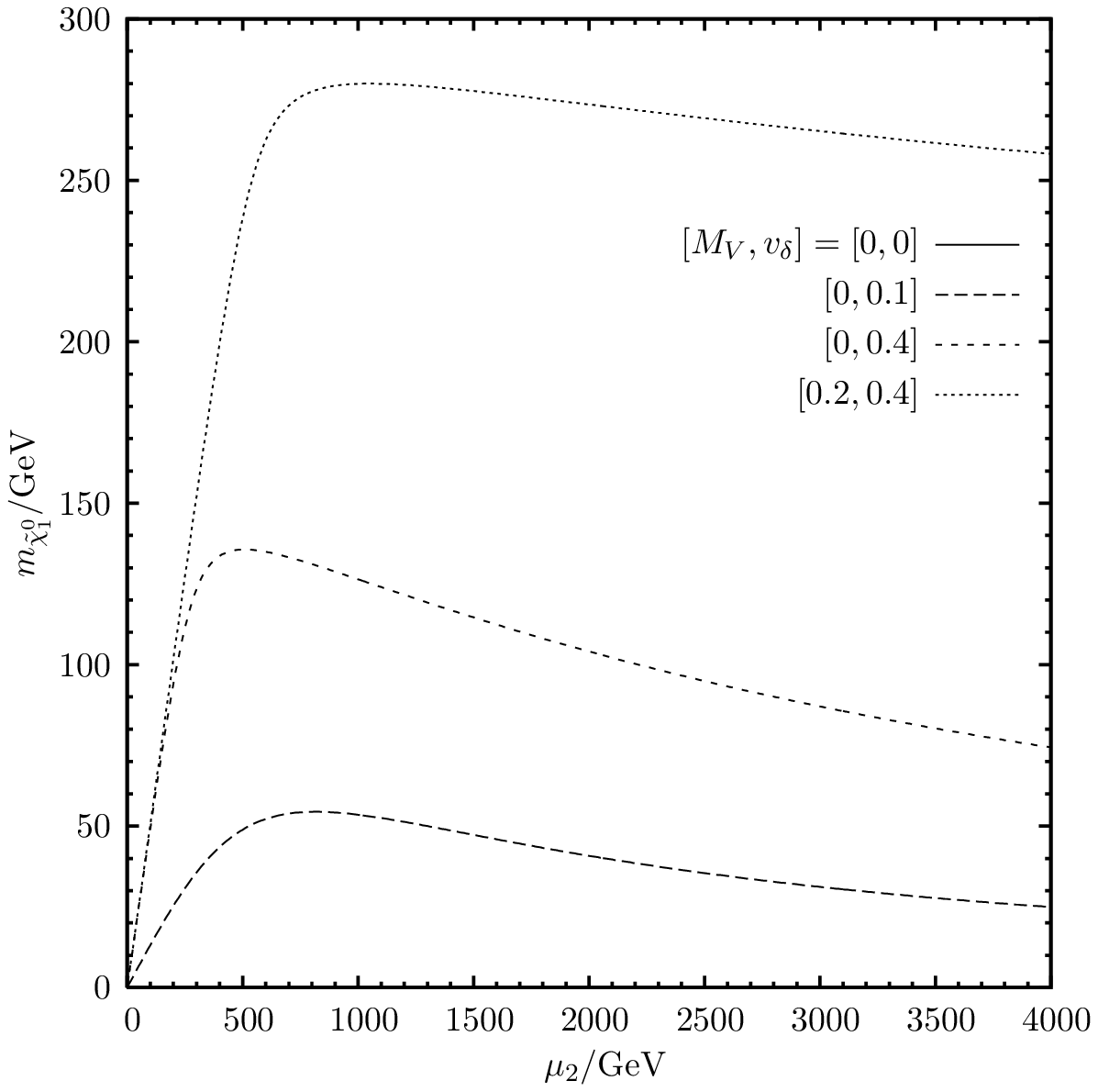}} }
\vskip -0.1in
\caption{{\it On the left panel, the difference between bino and higgsino
compositions of the lightest
neutralino state as a function of $\mu_2$ for various
$(M_V,v_{\delta})$ values in TeV, for $M_L=M_R=600$ GeV, $\mu_1=500$ GeV,
$v_{\Delta}=1.5$ TeV, and $\tan\beta=2$. On the right panel, the mass
of the lightest neutralino as a function of $\mu_2$ for the same
parameter values.}}
\label{fig:BinoHiggsino}
\end{figure}

In the left panel of Fig.~\ref{fig:BinoHiggsino}, the difference
between the bino and higgsino compositions of the lightest
neutralino state is shown as a function of $\mu_2$ by choosing
$(M_V,v_{\delta})=(0,0),(0,100),(0,400)$, and $(200,400)$ GeV. The
rest of the parameters are fixed as $M_L=M_R=600$ GeV, $\mu_1=500$
GeV, $v_{\Delta}=1.5$ TeV, and $\tan\beta=2$. In the right panel,
the mass of the lightest neutralino is given as a function of
$\mu_2$ for the same parameter values. As seen from the figure the
lightest state is pure higgsino $\tilde{\delta}_R^0$ for very
small $\mu_2$ values, regardless of the values of
$(M_V,v_{\delta})$, as long as they are smaller than $500$ GeV.
The ratio is more sensitive to the vev of the right-higgsino
($v_{\delta}$) than to the $U(1)_{B-L}$ gaugino paramater $M_V$.
At large $\mu_2$ the state becomes pure bino from pure higgsino
and its mass shows a strong dependence  on $\mu_2$. So, one can
consider these two limiting cases as realizations of the scenarios
(1) and (3), mentioned above. From the right panel, one can
conclude that, except for very small values of $\mu_2$, the cases
where $M_V$ and $v_\delta$ are larger than 200 GeV predict a
lightest neutralino  with a mass in the range of 200-300 GeV.
Otherwise its mass remains less than 150 GeV.


\begin{figure}[htb]
\vspace{-0.4in}
    \centerline{\hspace*{-0.2cm} \epsfxsize 3.57in
{\epsfbox{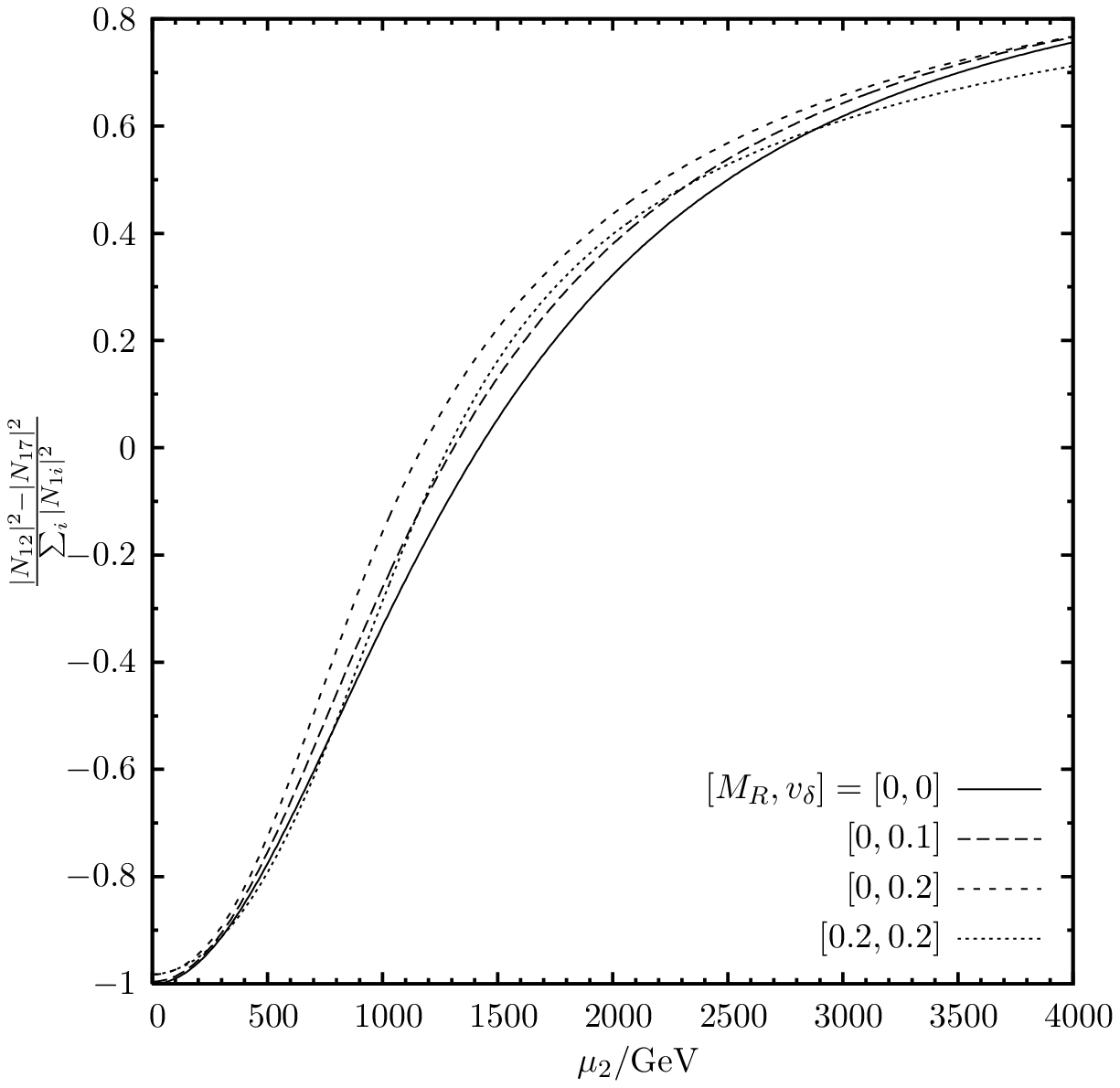}} \hspace{-0.65cm} \epsfxsize
3.5in {\epsfbox{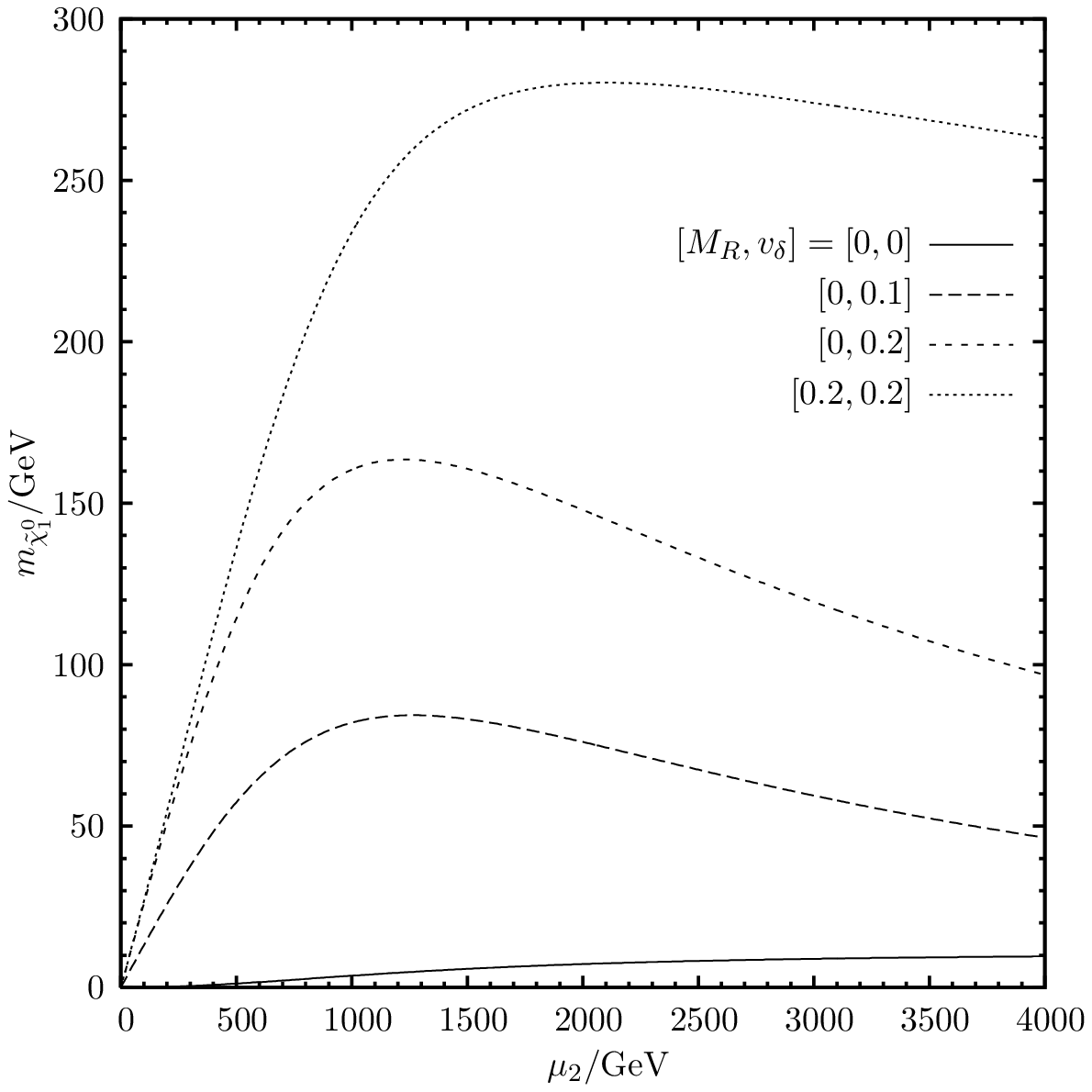}}}
 \vspace*{0.2cm}
\centerline{\epsfxsize 3.5in
{\epsfbox{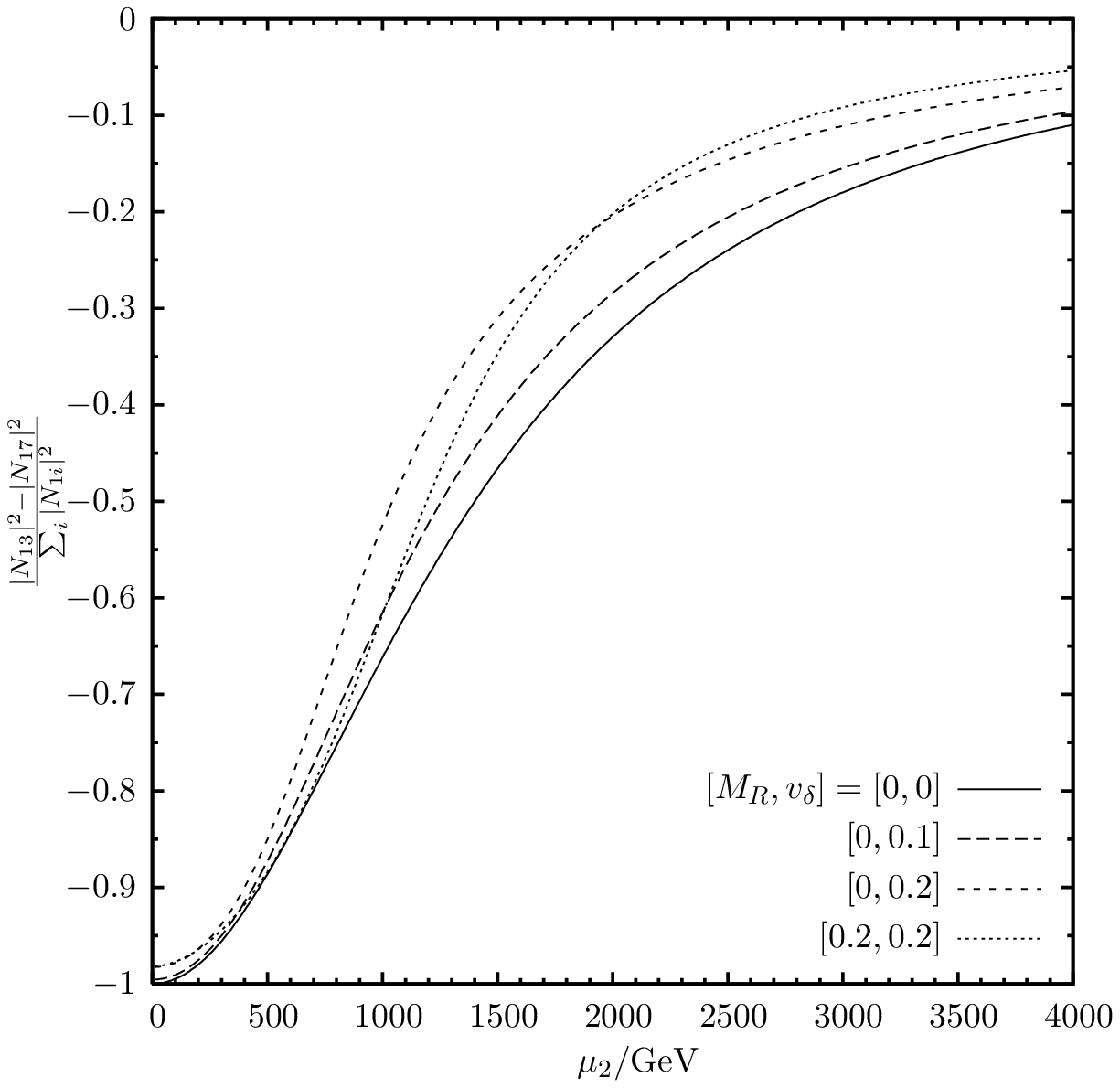}}} 
\vskip -0.1in
\caption{{\it  On the left panel, the difference between
right-wino and higgsino compositions of the lightest neutralino
state as a function of $\mu_2$ for various $(M_R,v_{\delta})$
values in TeV, for $M_L=M_V=600$ GeV, $\mu_1=500$ GeV,
$v_{\Delta}=1.5$ TeV, and $\tan\beta=2$. On the right panel, the
mass of the lightest neutralino as a function of $\mu_2$ for the
same parameter values. The third diagram shows the difference
between  bino and higgsino compositions of the the lightest
state.}}
    \label{fig:WinoHiggsino}
\end{figure}

{\it Next we look at Scenarios (2) and (3). Here we want the
right-wino to be the lightest neutralino, so we take $M_L$, $M_V$
and $\mu_1$ large, and $v_{\delta}$ to be small. Varying $\mu_2$
will take us from a mostly right-wino lightest state to a mostly
$\tilde{\delta}_R^0$ lightest state. }

In Fig.~\ref{fig:WinoHiggsino}, we plot the difference between the
bino and the higgsino compositions of the lightest neutralino state as a
function of $\mu_2$ for various $(M_R,v_{\delta})$ values, for
$M_L=M_V=600$ GeV, $\mu_1=500$ GeV, $v_{\Delta}=1.5$ TeV, and
$\tan\beta=2$. On the right panel, the mass of the lightest
neutralino is shown as a function of $\mu_2$ for the same parameter
values. The third diagram shows the bino composition of the
lightest state. This case exhibits a very similar dependence on $\mu_2$
as the bino-higgsino case. The curves for various
$(M_R,v_{\delta})$ pair indicate that the ratio is insensitive to
their chosen values, as long as they are assumed to be less than
$200$ GeV. The  bino composition becomes significant only around
$\mu_2=1$ TeV, where the right-wino and higgsino mix almost equally,
and it is negligible as $\mu_2$ becomes larger. The contribution of the bino with
respect to that of the right-wino is significant for very small $\mu_2$ values ($\sim 0$ GeV) but
such small values are excluded from the lower bound on the mass of
the lightest chargino, which is around $90$ TeV. The mass of the lightest state
is less than 150 GeV for $v_\delta\le 200$ GeV if $M_R$ is very small, or for
$(M_R,v_\delta)\le (50,50)$ GeV. Otherwise, it is larger than 200 GeV for $\mu_2$
larger than $500$ GeV. While the composition of the state is insensitive to $M_R$
and $v_{\delta}$,  the mass of the lightest neutralino is sensitive
to both. Here one can recover
Scenarios (2) and (3) in the limiting cases and the mixed state
($\vert{\lambda}^0_R\rangle+\vert\tilde{\delta}^0_R\rangle$) in between.

{\it Finally, we look at realizations of Scenarios (4), (5), and
(6). Here we are interested in mixtures of bino-(non MSSM)
higgsino, right-wino-(non MSSM) higgsino, and right-wino-MSSM
higgsino. We need to decouple the MSSM particles for scenarios (4)
and (5), so we take $M_L$ and $\mu_1$ to be large. In addition we
take $v_{\Delta}=v_{\delta}$ (as opposed to the previous
scenarios, where we took one of them small). We vary $M_V, \,M_R$
and $\mu_2$ to go from one case to another.}

The realization of the last three  scenarios discussed above
requires some radical changes in the parameter set and has a distinct
structure. We first decouple the  $\tilde{\phi}_{12}^0$ and
$\tilde{\phi}^0_{21}$ fields. Then, without loss of generality, we rotate
the basis
\begin{eqnarray}
\left\{\vert{\lambda}^0_L\rangle,\vert{\lambda}^0_R\rangle,\vert{\lambda}_V\rangle,\vert\tilde{\phi}^0_{11}\rangle,\vert\tilde{\phi}^0_{22}\rangle,\vert\tilde{\Delta}^0_R\rangle,\vert\tilde{\delta}^0_R\rangle\right\}
\to
\left\{\vert{\lambda}^0_L\rangle,\vert{\lambda}^0_R\rangle,\vert{\lambda}_V\rangle,\vert\tilde{\phi}^0_+\rangle,\vert\tilde{\phi}^0_-\rangle,\vert\tilde{\eta}^0_+\rangle,\vert\tilde{\eta}^0_-\rangle\right\}
\end{eqnarray}
such that
\begin{eqnarray}
\vert\tilde{\phi}^0_\pm\rangle\equiv
\frac{\vert\tilde{\phi}^0_{11}\rangle\pm
\vert\tilde{\phi}^0_{22}\rangle}{\sqrt{2}},\;\;\;\vert\tilde{\eta}^0_\pm\rangle\equiv
\frac{\vert\tilde{\Delta}^0_R\rangle \pm
\vert\tilde{\delta}^0_R\rangle}{\sqrt{2}}.
\end{eqnarray}
In this new basis the mass matrix $Y$ in Eq.~(\ref{Ymatrix})
becomes
\begin{eqnarray}
\displaystyle
Y^\prime=\left(
\begin{array}{c|ccc|ccc}
  & -i \lambda_L^0 & -i \lambda_R^0 & -i
\lambda_V & \tilde{\phi}_{+}^0 &
\tilde{\phi}_{-}^0 & \tilde{\eta}_{-}^0 \\
\hline
-i \lambda_L^0 &
M_L & 0 & 0 &  \frac{g_L}{2} \kappa^- &
\frac{g_L}{2}
\kappa^+&0 \\
-i \lambda_R^0 & 0 & M_R  &
0 & -\frac{g_R}{2} \kappa^- &
-\frac{g_R}{2}
\kappa^+ & 2g_R v_R\\
-i \lambda_V & 0 & 0
& M_V & 0 & 0 & -2g_V v_R\\
\hline
\tilde{\phi}_{+}^0 & \frac{g_L}{2} \kappa^- &
-\frac{g_R}{2} \kappa^- & 0 &-\mu_1& 0
& 0\\
  \tilde{\phi}_-^0 & \frac{g_L}{2}
\kappa^+ & -\frac{g_R}{2} \kappa^+ & 0 & 0 &
\mu_1 & 0\\



  \tilde{\eta}_-^0
&  0 &2 g_R v_R & -2 g_V v_R & 0 & 0 &\mu_2
\end{array}
\right ),
\end{eqnarray}
where we have assumed $v_{\Delta}=v_{\delta}\equiv v_R$, and
further defined $\kappa^\pm=\kappa_1\pm\kappa_2$. Under this
assumption the rotated state $\tilde{\eta}_+^0$ decouples and only
$\tilde{\eta}_-^0$ remains, which allows us to analyze scenarios
(4), (5) and (6) more conveniently.

\begin{figure}[htb]
\vspace{-0.4in}
    \centerline{\hspace*{-0.2cm} \epsfxsize 3.45in
{\epsfbox{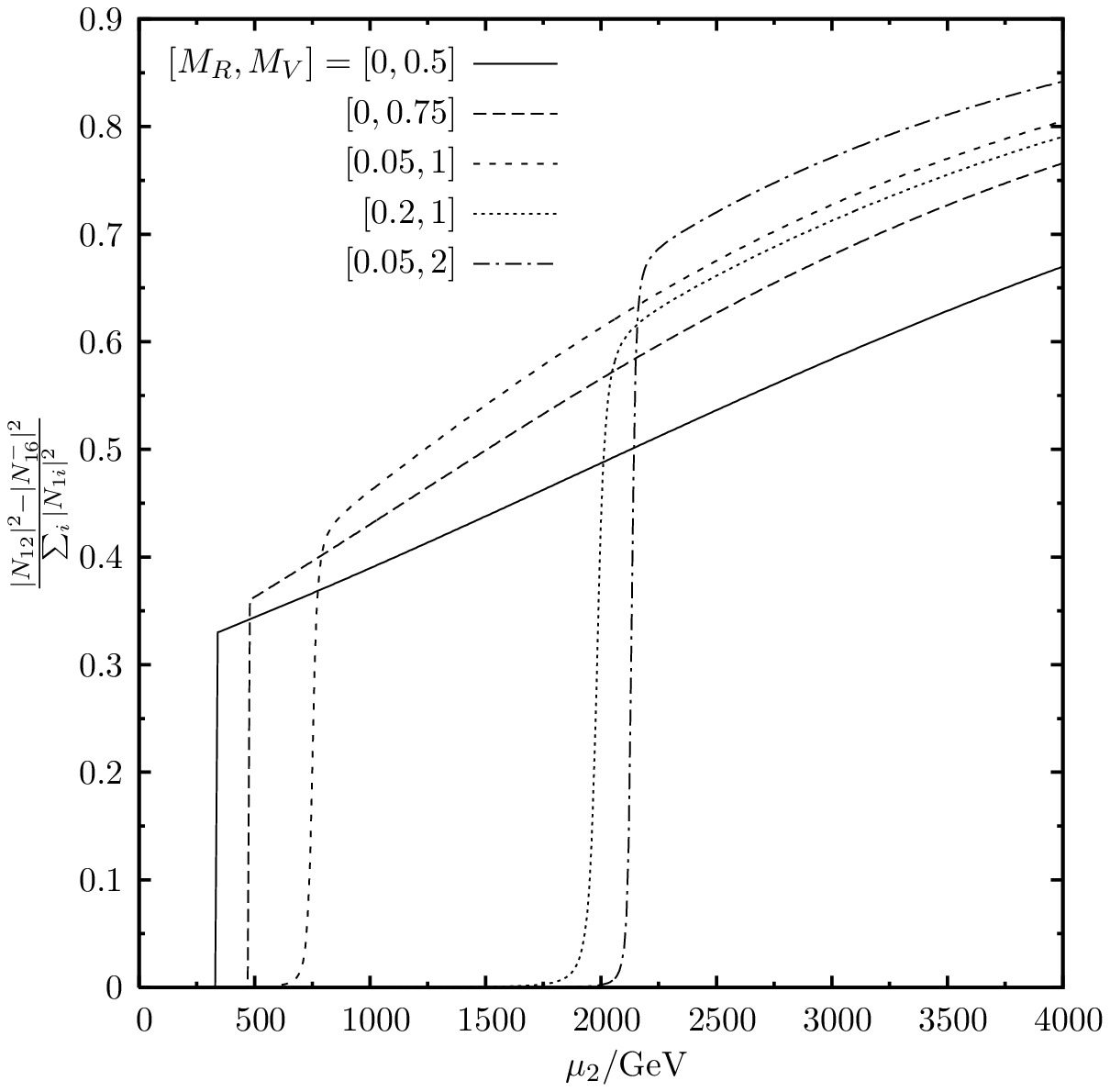}} \hspace{-0.65cm}
\epsfxsize 3.5in
{\epsfbox{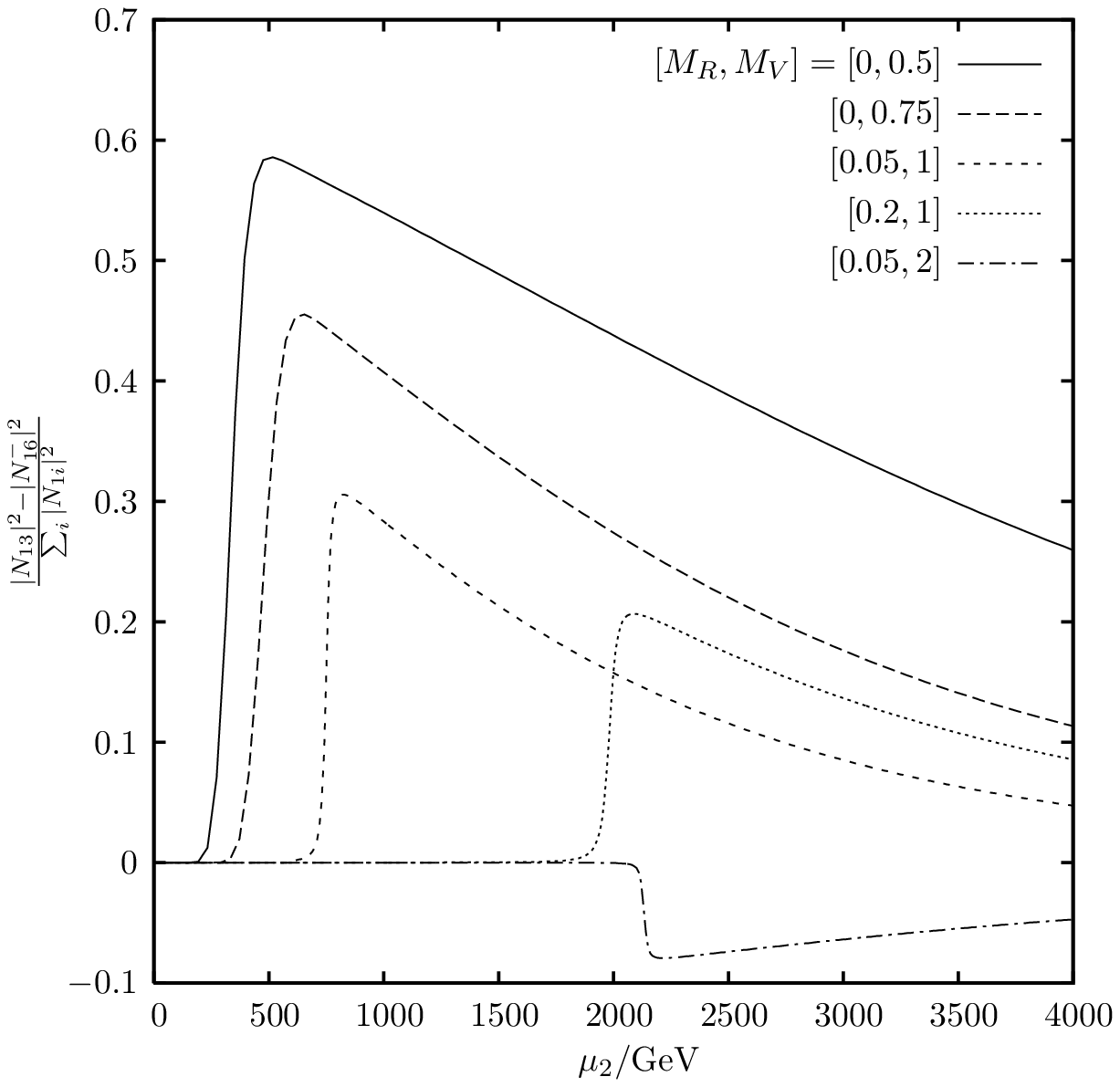}}}
\vspace*{0.1cm}

\centerline{\epsfxsize
3.5in {\epsfbox{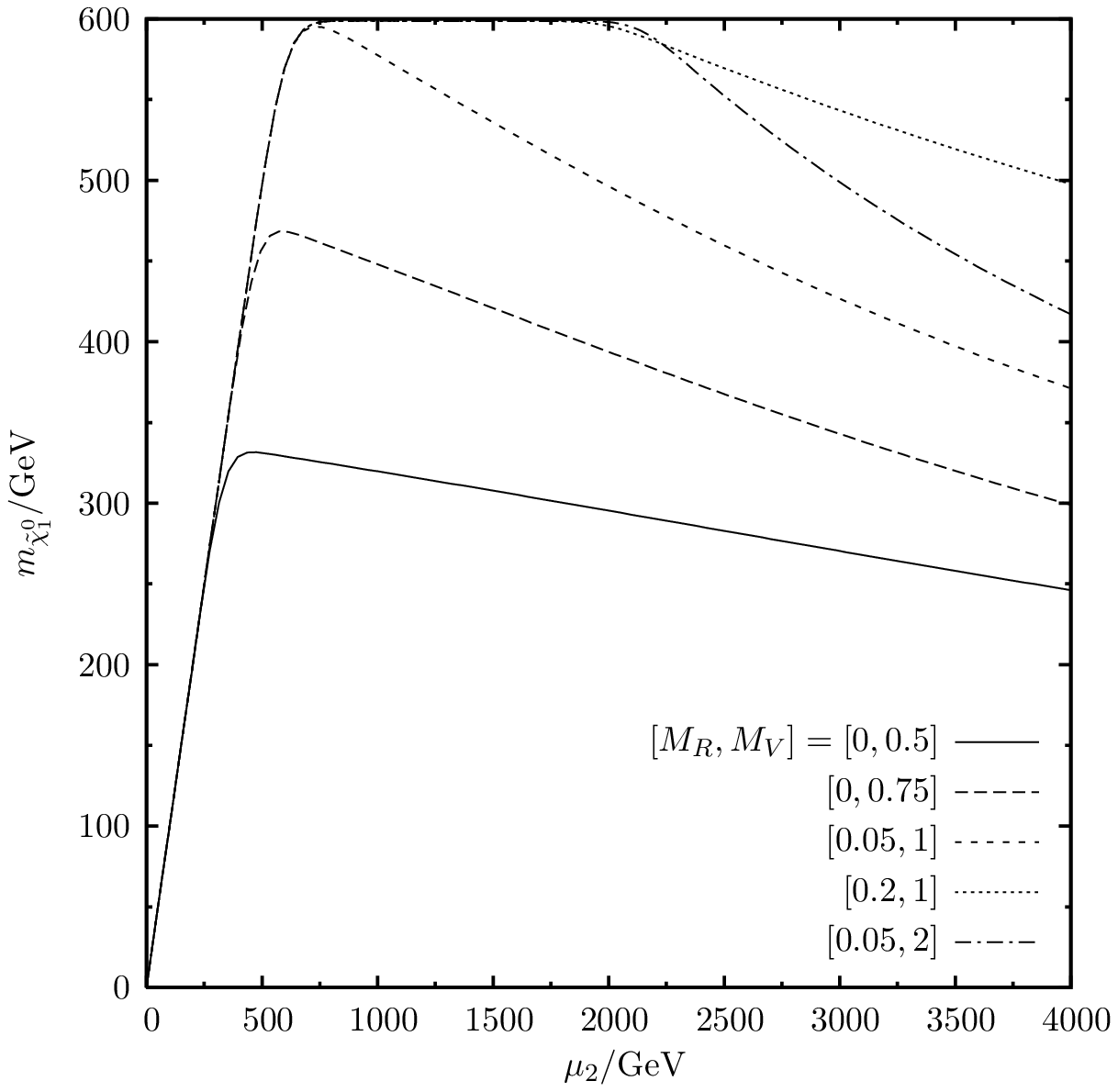}}}
\vskip -0.3in
 \caption{{\it On the left panel, the difference between
right-wino and rotated non-MSSM higgsino
compositions of the
lightest neutralino state as a function of $\mu_2$ for various
$(M_R,M_V)$ values in TeV, for $M_L=600$ GeV, $\mu_1=5$ TeV,
$v_{\Delta}=v_{\delta}=1$ TeV, and $\tan\beta=2$. On the right
panel, the difference between the
bino and the rotated non-MSSM higgsino
compositions as functions of $\mu_2$ for the same
parameter values. The third diagram shows the mass of the lightest
neutralino as a function of $\mu_2$.}}
 \label{fig:BinoWinoHiggsino}
\end{figure}

Fig.~\ref{fig:BinoWinoHiggsino} shows the difference between  the
right-wino (bino) and the rotated non-MSSM higgsino
($\tilde{\eta}_-^0$) compositions of the lightest neutralino state
as functions of $\mu_2$ by choosing various $(M_R,\,M_V)$ values
in the left (right) panel. The rest of the parameters are taken as
$M_L=600$ GeV, $\mu_1=5$ TeV, $v_{\Delta}=v_{\delta}\equiv v_R=1$
TeV, and $\tan\beta=2$. Here $|N_{16}^-|^2$ is the amount of the
rotated higgsino field $\tilde{\eta}_-^0$ in the lightest
neutralino state. One can divide the discussion into two
regions--one with small $M_R\le 50{\rm GeV}$ and $M_V\le 1$ GeV,
and the other with larger $M_R$ and $M_V$. In the first part,
neither the right-wino nor  the bino is the lightest neutralino
till $\mu_2\sim 500$ GeV. The decoupled state $\tilde{\eta}_+^0$
is the lightest state (which can not be seen from the
figures\footnote{Note that the horizontal line passing through
zero does not always mean equal mixing.  One could obtain
such a result for vanishing individual contributions as well, which is 
indeed the case here.}).
For $\mu_2 > 500$ GeV,  the lightest state is mainly a right-wino
or a right-wino-higgsino mixture. Otherwise,  for intermediate
values of $\mu_2$, it is a mixture of right-wino, bino, and
higgsino $\tilde{\eta}_-^0$. For the second part where $M_R > 50$
GeV and $M_V > 1000$ GeV, the curves are horizontal lines passing
through zero for both graphs until $\mu_2\sim 2$ TeV. In that
region of the parameter space the lightest state is pure
left-wino. The right-wino-higgsino mixture is possible as the
lightest neutralino state only after that point. The bino-higgsino
lightest state is not realized in this region. From the third
diagram in Fig.~\ref{fig:BinoWinoHiggsino}, one can say that
within the range considered for $\mu_2$, the mass of the lightest
state is greater than 300 GeV for $\mu_2\ge 500$ GeV. So, the
scenarios (4) and (5) require the lightest state to be rather
heavy unless $\mu_2$ is too large. This takes us to pure
right-wino scenario. Including all the possible model constraints
in the next section will restrict our parameter space further.

\begin{figure}[htb]
\vspace{-0.05in}
    \centerline{\hspace*{-0.2cm} \epsfxsize 3.72in
{\epsfbox{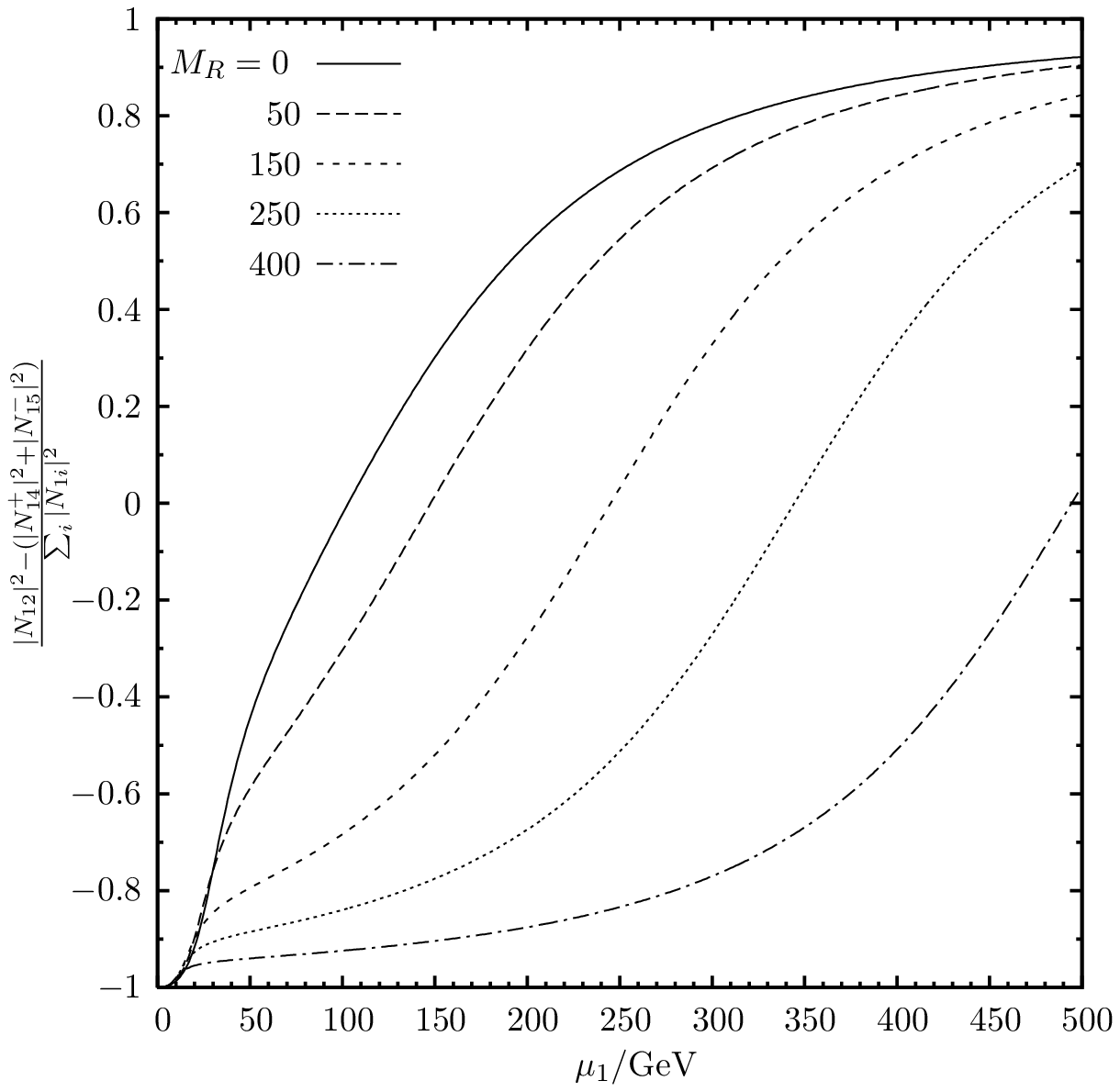}} \hspace{-0.65cm}
\epsfxsize 3.53in
{\epsfbox{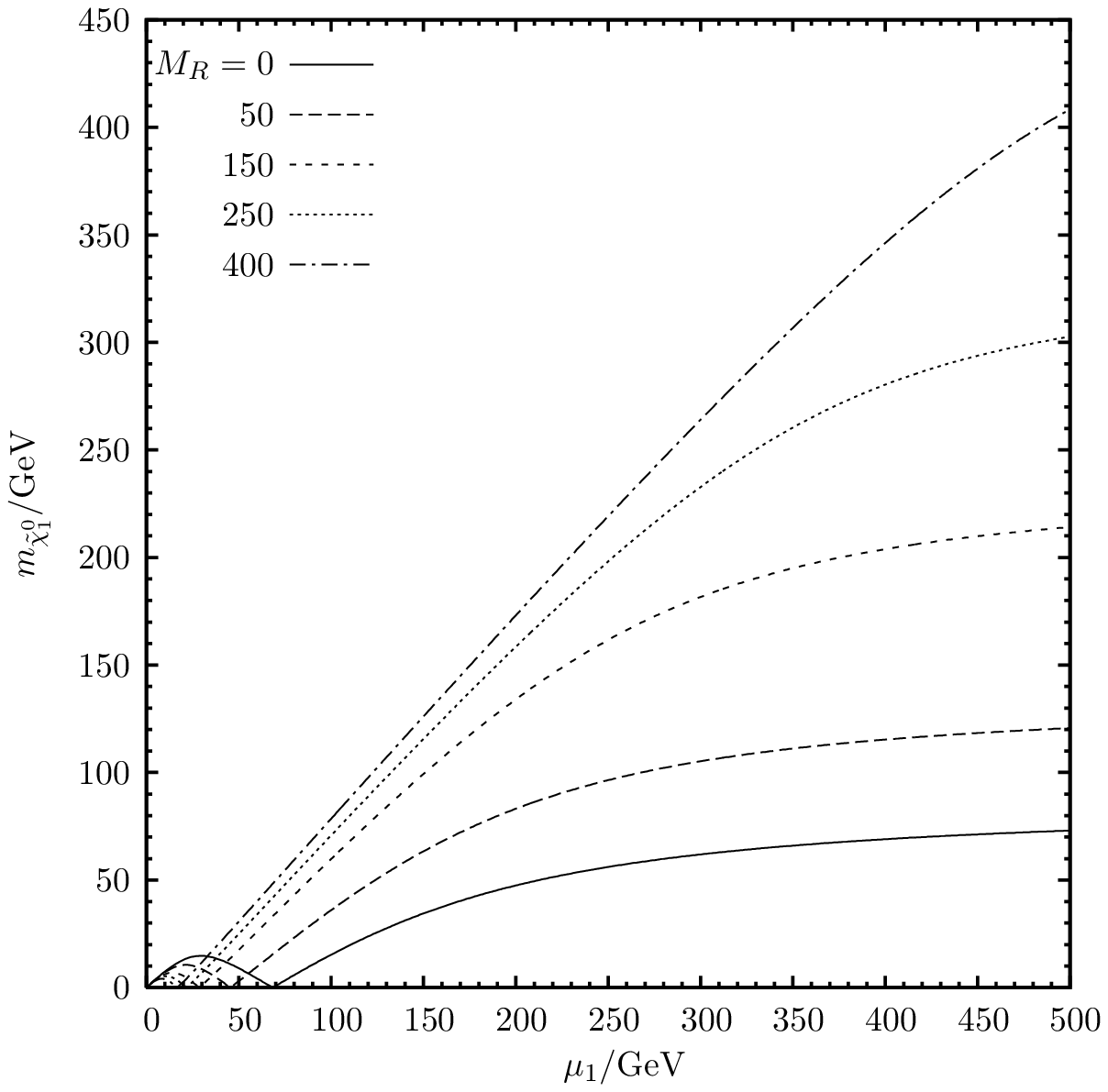}}}
\vskip -0.1in
 \caption{{\it On the left panel, the difference between
right-wino and rotated MSSM higgsino
compositions of the
lightest neutralino state as functions of $\mu_1$ at various
$M_R$ values in GeV, for $M_L=M_V=1$ TeV, $\mu_2=20$ TeV,
$v_{\Delta}=v_{\delta}=1$ TeV, and $\tan\beta=2$. On the right
panel, the mass of the lightest
neutralino as a function of $\mu_1$ for the same parameter values.}}
 \label{fig:WinoMSSMHiggsino}
\end{figure}

In Fig.~\ref{fig:WinoMSSMHiggsino}, the difference between  the
composition of the lightest neutralino as a right-wino, or a
MSSM-higgsino is shown on the left panel as a function of $M_R$
for various $\mu_1$ values in the range  $0$ to $500$ GeV. The
other parameters are chosen as  $M_L=M_V=1$ TeV, $\mu_2=20$ TeV,
$v_{\Delta}=v_{\delta}=1$ TeV, and $\tan\beta=2$. The ratio is
sensitive to both $M_R$ and $\mu_1$. For values of $\mu_1$ smaller
than $25$ GeV, the lightest state is pure MSSM-higgsino and for
values larger than $450$ GeV the state almost is pure right-wino,
if $M_R$ is equal to, or less than, $150$ GeV. Evenly mixed states
are obtained for intermediate values of $\mu_1$. As $M_R$ becomes
equal to, or larger than, $400$ GeV, the lightest state remains
pure right-wino for the most part of the $\mu_1$ range. As
suggested by  the right panel, the mass of the lightest state is
very sensitive to the parameter $M_R$ and can be quite large for
large $M_R$ and $\mu_1$ values.

This completes analyses of possible LSP candidates in LRSUSY for
certain patterns of the model parameters. The figures illustrate
nature and purity of the lightest neutralino states as a function
of various gaugino masses and Higgs vevs. In the next section we
will perform a detailed analysis of the relic abundance of LSP and
its ability to explain the CDM in the universe.

\section{Relic abundance analysis of CDM}
\label{relicdensity}
In this section we calculate the relic density of cold dark matter within the LRSUSY framework.
Before presenting our results we give some details of the procedure followed.

The time evolution of the number density $n_i$ for a relic
particle is given by the Boltzmann equation (taken within
spatially flat Friedman-Robertson-Walker background). Furthermore, a single Boltzmann
equation can be defined \cite{Griest:1990kh} for the total number
density, $n\equiv \sum_i n_i$
\begin{eqnarray}
\dot{n}+3h n =-\langle \sigma_{eff} v\rangle \left[n^2-n_{\rm eq}^2\right],
\label{BE}
\end{eqnarray}
where $h$ is the Hubble parameter, $v$ is the relative velocity of
the annihilating particles, and $n_{\rm eq}$ is the number density
corresponding to the sum of each species number density at thermal
equilibrium (that is, the density in the early universe). Here
$\sigma_{eff}$ is the properly averaged effective cross section of
the CDM candidate into ordinary particles (i.e., SM particles
including the Higgs bosons) and will be defined below. Clearly,
``$\langle\,\, \rangle$" stands for the thermal average.

In this study we concentrate on regions of the parameter space
where the coannihilation effects are not significant. The
relativistic thermally-averaged cross section times relative
velocity reads as \cite{Gondolo:1990dk}
\begin{eqnarray}
\displaystyle
\langle \sigma_{eff} v\rangle(x) = \frac{\int_{4m_\chi^2}^{\infty}ds\,\sigma(s)\,\sqrt{s}\,(s/4-m_\chi^2)\,K_1\left(\sqrt{s}/(m_\chi x)\right)}
{2 m_\chi^5\, x\, K_2\left(1/x\right)^2},
\label{sigmav}
\end{eqnarray}
where $m_\chi$ is the mass of the annihilating particle, the
lightest neutralino in our consideration; $K_1$ and $K_2$ are
modified Bessel functions; $x\equiv T/m_\chi$ is the dimensionless
temperature parameter; $\sqrt{s}$ is the center of mass energy of
the annihilating neutralino pair; and $\sigma(s)$ is the cross
section for the annihilation reaction $\chi_1^0 \bar{\chi}_1^0\to
{X_{\rm SM}}$, where $X_{\rm SM}$ represents all the allowed
two-body SM final states.  We will not approximate Eq.~(\ref{sigmav})
by expanding in $v$ in a Taylor series, as this approximation
might not always be safe,  especially for cases with  fast varying
integrands.

In general, there is no analytical solution to the Boltzmann
equation, Eq.~(\ref{BE}), which is a Riccati-type equation. Thus a
numerical approach is required. There exist several different ways
to proceed in the literature  \cite{Belanger:2004yn,Gondolo:1990dk}.
Here we define a freeze-out temperature parameter $x_F\equiv
T_F/m_\chi$, which we use as an approximate solution to
Eq.~(\ref{BE}) \cite{Baer:2003jb} as
\begin{eqnarray}
\displaystyle
n_\chi(T_0)\sim\frac{1}{m_\chi M_{Pl}}\sqrt{\frac{4\pi^3 g*}{45}}\left(\frac{T_\chi}{T_\gamma}\right)^3 T_\gamma^3\frac{1}{\int_{x_0}^{x_F} dx \langle\sigma_{eff}v\rangle\!\left(x\right)},
\end{eqnarray}
where $M_{Pl}$ is the Planck mass, $T_0$ is today's temperature
($\sim 2.742 K^\circ$), so that $x_0$ can be approximated as zero;
$T_\chi(T_\gamma)$ is the present temperature of the neutralino
(Cosmic Microwave Background), and $g*$ represents the SM
effective number of degrees of freedom at the freeze-out
temperature ($g*\sim 81$ is used). Note that using an approximate
solution of the Boltzmann equation instead of solving it
numerically will introduce an uncertainity of up to $10\%$ into
our results.

The freeze-out temperature parameter $x_F$ can be obtained from the following transcendental equation
\begin{eqnarray}
\frac{1}{x_F}={\rm log}\left[\frac{m_\chi M_{Pl}}{2\pi^3}\sqrt{\frac{45}{2g*}}\sqrt{x_F}\,\langle \sigma_{eff} v \rangle\!\left(x_F\right)\right]
\end{eqnarray}
which can be solved iteratively. As a starting point we need to choose a value for $x_F$. We used 1/25 in our calculations.
The range for $x_F$ changes between 1/25 to 1/15. Finally, the relic density of neutralinos at present time is defined as
\begin{eqnarray}
\Omega_\chi h^2 = \frac{\rho(T_0)}{\rho_c}
\end{eqnarray}
where $\rho(T_0)=m_\chi n_\chi(T_0)$ is the density of the
neutralino, and $\rho_c=3h^2M_{Pl}^2/8\pi=8.098\times 10^{-47}\,
{\rm GeV^4}$ is the critical density of the universe. On the left
hand side of the equation, $h$ is the normalized Hubble expansion
rate and its  value today is 0.71. Thus from the central value of
$\Omega_\chi h^2$, $\Omega_\chi$ is found to be about $22\%$.
\begin{figure}[htb]
\vspace{-0.05in}
    \centerline{ \epsfxsize 5.5in
{\epsfbox{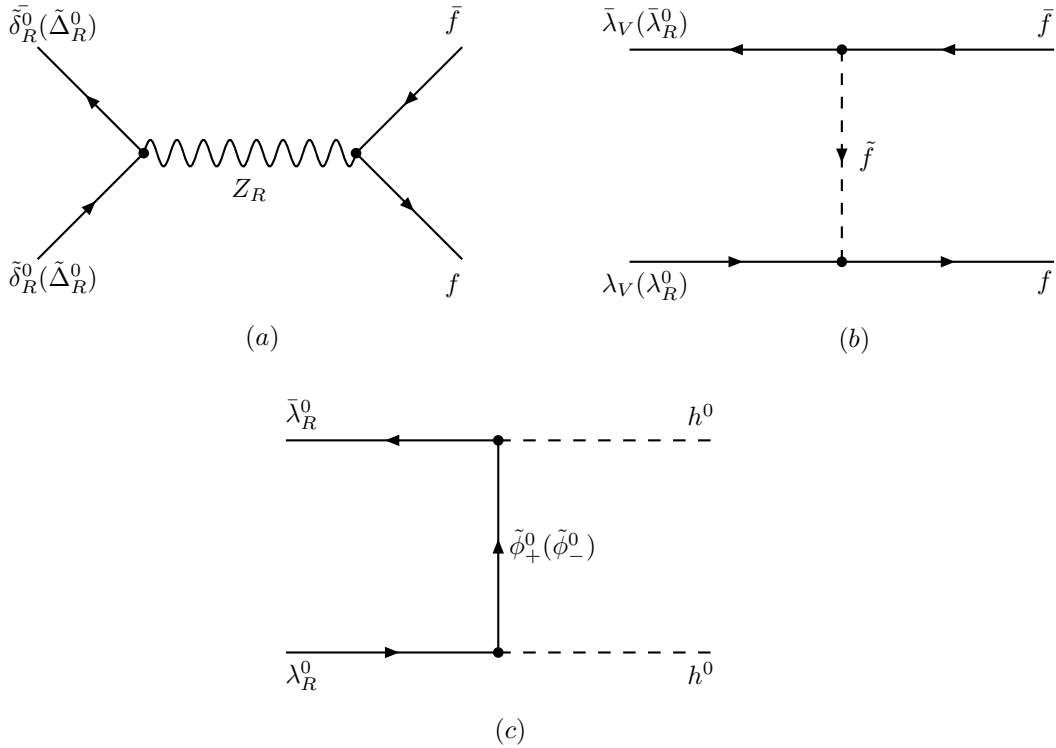}}}
\vskip -0.1in
 \caption{{\it The annihilation Feynman diagrams contributing $\tilde{\chi}_1^0\bar{\chi}_1^0\to {\rm X_{SM}}$ in the LRSUSY model. The MSSM contributions and $u-$channel diagrams are not shown here. Contributions are given in the basis of pure states whose mixtures can be easily constructed. Here $f$ represent all fermions, and $\tilde{f}$ all sfermions, $h^0$ is the SM-like Higgs particle, and  $\tilde{\phi}_+^0, \tilde{\phi}_-^0$ are the rotated higgsinos fields defined in the previous section.}}
 \label{fig:feyndiag}
\end{figure}

To obtain our final results, a three dimensional integration needs
to be carried out numerically. The three parts are: an integration
over the Mendelstam variable $t$ to compute $\sigma(s)$ for each
subprocess in the kinematically allowed region, an integration
over the center of mass energy squared $s$ from the threshold to
practically infinity, and finally an integration over $x$ to
compute the thermal averaging from the freeze-out temperature to
today's ($x_0$), which we approximate to zero.
\texttt{Mathematica} is used for the computation and the matrix
element calculations have been carried out with  \texttt{FeynCalc}
\cite{Mertig:1990wm}.

The Feynman diagrams contributing to the annihilation cross
section are given in Fig.~\ref{fig:feyndiag}.  We analyze here
pure state contributions only, while any mixed case scenario can
be calculated in a straightforward manner. Since our analysis
concentrates on the non-MSSM scenarios, the MSSM contributions are
not shown here. Depending of the center of mass energy available,
the resonance problem in $s$-channel is handled using the
Briet-Wigner prescription.

The mass of the $Z_R$ boson in diagram $(a)$ is
 $M_{Z_R}=\displaystyle{\frac {g \cos
\theta_W}{\sqrt{\cos 2\theta_W}}
(v_{\delta}^2+v_{\Delta}^2)^{1/2}}$ \cite{Frank:2003un}. In diagram $(b)$ we sum over
all left-- and right--handed quarks and leptons in the final
state. The sfermion $\tilde f$ has a (mass)$^2$ given
by, for U-type squarks and sneutrinos:
\begin{equation}
{\cal M}_{U_k}^2= \left( \begin{array}{cc}
                             M_{\rm SUSY}^2+M_{Z_L}^2(T_u^3-Q_u \sin^2 \theta_W) \cos 2
\beta & m_{u_k} (A-\mu \cot \beta) \\
                            m_{u_k} (A-\mu \cot \beta) &
M_{\rm SUSY}^2+M_{Z_L}^2 Q_u \sin^2 \theta_W \cos 2 \beta
                        \end{array}
                 \right),
\end{equation}
and for D-type squarks and sleptons
\begin{equation}
{\cal M}_{D_k}^2= \left( \begin{array}{cc}
                             M_{\rm SUSY}^2+M_{Z_L}^2(T_d^3-Q_d \sin^2 \theta_W) \cos 2
\beta & m_{d_k} (A-\mu \tan \beta)\\
                           m_{d_k} (A-\mu \tan \beta) &
M_{\rm SUSY}^2+M_{Z_L}^2 Q_d \sin^2 \theta_W \cos2 \beta
                        \end{array}
                 \right).
\end{equation}
Here $M_{\rm SUSY}$ represents the  universal scalar mass, $A$ and
$\mu$ the trilinear and bilinear scalar parameters, respectively,
$m_{u,d}$ are quark masses and the index $k$ labels generations.
It is customary to restrict the supersymmetric parameter space by
drawing contours in the $M_{1/2}-M_{\rm SUSY}$ plane, where
$M_{1/2}$ is the relevant gaugino or higgsino mass parameter and $M_{\rm
SUSY}$ the relevant scalar mass. When the bino or right-wino are
the LSP, we take their masses to be the significant gaugino mass. In
diagram $(c)$ the final state $h^0$ is the MSSM-like lightest
Higgs boson, which we constrain to have a mass $M_{h^0}=115$ GeV.
In LRSUSY, its (mass)$^2$ is given by the lowest eigenvalue of the
matrix \cite{Frank:2003un}:
\begin{equation}
{\cal M}_{\phi_{22}, \phi_{11}}^2= \left( \begin{array}{cc}
                             m_{\phi_1\phi_2}^2 \cot \beta-g^2v^2 & -m_{\phi_1\phi_2}^2\\
                            -m_{\phi_1\phi_2}^2 &
m_{\phi_1\phi_2}^2 \tan \beta+g^2v^2
                        \end{array}
                 \right).
\end{equation}
where $v^2=v_{\delta}^2+v_{\Delta}^2+\kappa_1^2+\kappa_2^2$ and
$m_{\phi_1\phi_2}$ is the Higgs mass parameter. Note that while
the cross sections for the decay of $\lambda_V,\, \lambda_R^0$
depend on $M_{\rm SUSY}$, the one for the $\tilde \delta_R^0$
higgsino does not; we must find another relevant parameter for
that case.

\section{Numerical Analysis}

Based on our classification of some possible mass scenarios in
Section~\ref{classification}, we analyze the first three
mass scenarios with the lightest neutralino as a pure bino,
right-wino, or higgsino state. The feasibility of the last three
scenarios, assuming mixed gaugino-higgsino states, can be analyzed in
the light of the pure state predictions.  Throughout the numerical
calculations, we have chosen the parameters of the model such that
the $\Delta \rho$ bound, taken as $\Delta\rho\le0.002$
\cite{Eidelman:2004wy}, is always satisfied.
\begin{figure}[htb]
\vspace{-0.05in}
    \centerline{\epsfxsize 5.5in
{\epsfbox{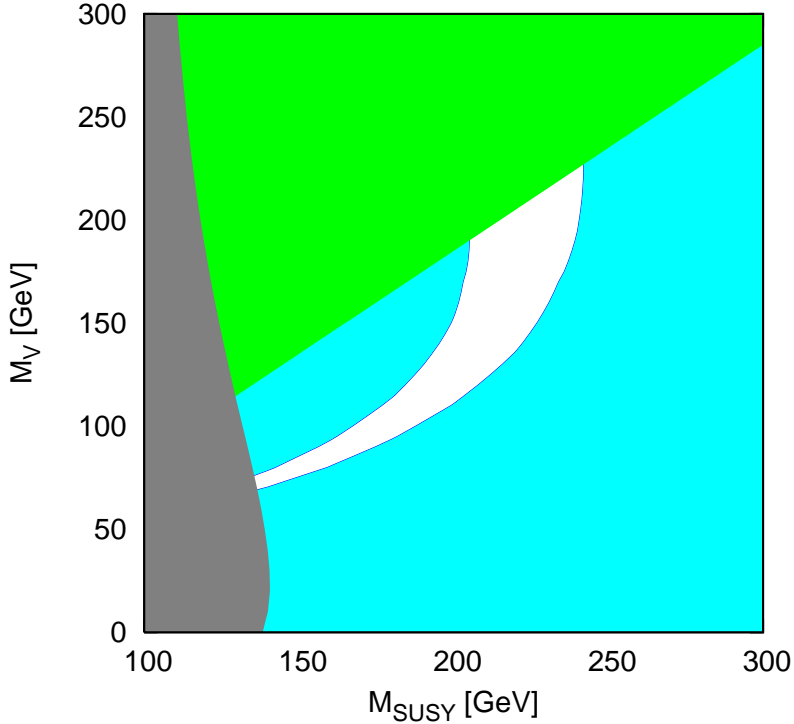}}}
\vskip -0.1in
 \caption{{\it The relic density of a pure bino state in $M_V-M_{\rm SUSY}$ plane for $\mu_1=500$ GeV,
$A=200$ GeV, and $\tan\beta=2$, in the scenario in which the bino is the lightest neutralino. The  upper triangular region in green is excluded by the lower bound requirement of
the lightest neutralino mass from LEP2 and by the lower bound on the lightest sfermion mass constraint. The light blue region is excluded
by WMAP measurements. The gray vertical strip on the left is excluded by $b\to s\gamma$.  The white strip agrees with the WMAP at $2\sigma$ level. The anomalous magnetic moment of the muon is satisfied within the entire region.}}
 \label{fig:omegaBino}
\end{figure}

Fig.~\ref{fig:omegaBino} shows the relic density $\Omega_\chi h^2$
of the neuralino as a pure bino state  in the $M_V-M_{\rm SUSY}$
plane for $\mu_1=500$ GeV, $A=200$ GeV, and $\tan\beta=2$. There
is no $s$-channel contribution and the $t$-channel fermion pair
production $\lambda_V\bar{\lambda}_V\stackrel{\tilde{f}}{\to}
f\bar{f},\,\,(f=u,c,t,d,b,s...)$ is the only channel available.\footnote{We 
recall that the bino here is the fermionic partner of
the $U(1)_{B-L}$ gauge boson and is different from the bino in the
MSSM.} Due to the $s$-wave supression, the fermion pair
production for heavy fermions is  dominant, but we include all
possible channels. Unlike the MSSM bino, the $B-L$ bino does not
couple to $W^+W^-$ pairs or to Higgs pairs. For the exchanged
sfermions, a flavor conserving scenario with a common scale
$M_{\rm SUSY}$ is assumed. In the figure, the upper triangular 
region in green is
excluded by the lower mass bound requirement on the lightest
neutralino from LEP2 and by the lower bound of the lightest
sfermion mass  determined mainly by $M_{\rm SUSY}$. (We calculated the masses of the sfermions and require that they are at least $15$ GeV higher than the mass of the bino to avoid co-annihilation channels). The light 
blue regions are excluded by the WMAP measurements. In addition, the 
gray vertical strip on the left is excluded by the branching ratio of 
$b\to s\gamma$
 upper bound, for which the 
formulas are given in the Appendix. We assumed that  
$2.4\times 10^{-4}\le BR(b\to s\gamma) \le 4.1\times 10^{-4}$ range is allowed. 
In numerical calculation, we used the following set of parameter values, consistent with 
pure bino scenario:
$M_L=M_R=3$ TeV, $m_{\tilde{g}}=400$ GeV, $A=200$ GeV, $\mu_1=500$ GeV, 
$\mu_2=4$ TeV, $v_\delta=100$ GeV, and  $v_\Delta=1.5$ TeV.\footnote{We ensure at all times that the bino is not just the lightest neutralino, but a pure state.}   As seen from the figure 
$BR(b\to s\gamma)$ constraint excludes light $M_{\rm SUSY}$ ($\le 150$ GeV) region.
We also take into account of the anomalous magnetic moment of muon $a_\mu=(g-2)_\mu/2$, 
where the analytical formulas are given in the Appendix. The dominant chargino-sneutrino and neutralino-smuon loop contributions are included. 
In numerical computation, like $BR(b\to s\gamma)$, the same parameter set is assumed and the 
range for the deviation from the SM, $-0.53 \times 10^{-10} \le \Delta a_{\mu} \le 44.7 \times 
10^{-10}$, at 95\% CL is used based on $e^+e^-$ data (see \cite{Lazarides:2006jw} and references therein). 
$\Delta a_\mu$ depends mainly on the chosen values of $A$, $\mu_1$, $\tan\beta$, and $M_{\rm SUSY}$. We observe 
that it remains in the allowed range throughout the entire $(M_{\rm SUSY},M_V)$ interval considered, as long as $\mu_1\tan \beta-A \ge 0$ holds. 

Consequently, the
white strip is the only region satisfying the WMAP at $2\sigma$
level. A realization of such a scenario requires light sfermion
masses in the range $\sim 150-240$ GeV with an acceptable LSP mass
in the region $\sim 60-220$ GeV. In the light blue region on the right,
the relic density $\Omega_\chi h^2|_{\chi=\lambda_V}$ becomes
larger than experimentally allowed values, while in the left
painted region it is smaller than experimental requirements. This
scenario can only be considered feasible in frameworks where at
least there is at least one light sfermion, like  stop
$\tilde{t}$, or stau $\tilde{\tau}$.
\begin{figure}[htb]
\vspace{-0.05in}
    \centerline{ \epsfxsize 5.5in 
{\epsfbox{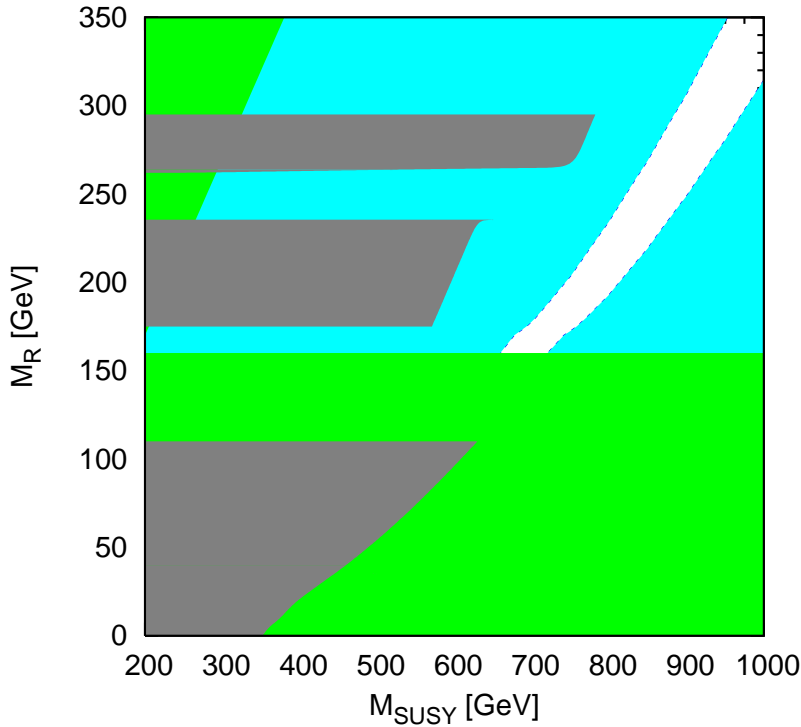}}}
\vskip -0.1in
 \caption{{\it The relic density of a pure right-wino state in $M_R-M_{\rm SUSY}$
plane for $\mu_1=1$ TeV, $A=200$ GeV, and $\tan\beta=2$ in the
scenario in which the right-wino is the lightest neutralino. The
left upper small triangular and the lower regtangular regions in green 
are excluded by the lower bound mass requirement
on the lightest neutralino from LEP2 and by the lower bound on the
lightest sfermion mass. The light blue region is excluded by WMAP
measurements. The grey region is excluded by $b\to s\gamma$. 
The white strip agrees with the WMAP at $2\sigma$
level. The anomalous magnetic moment of the muon is satisfied within the entire region.}}
 \label{fig:omegaRWino}
\end{figure}

Next, in Fig.~\ref{fig:omegaRWino} we show the relic density
$\Omega_\chi h^2$ of the neutralino as a pure right-wino state  in
the $M_R-M_{\rm SUSY}$ plane for $\mu_1=1$ TeV, $A=200$ GeV, and
$\tan\beta=2$. The right-wino as a neutralino is a completely new
possibility peculiar to the LRSUSY model. There is no $s$-channel
contribution and the possible annihilation contributions are
$\lambda_R^0\bar{\lambda}_R^0\stackrel{\tilde{f}}{\to}
f\bar{f},\,\,(f=u,c,t,d,b,s...)$ and
$\lambda_R^0\bar{\lambda}_R^0\stackrel {\tilde{\phi}_\pm^0}{\to}
h^0 h^0$, when kinematically allowed. An important feature in this
case is that there exists an additional condition -- the LEP lower
bound for the lightest chargino, which we take as $90$ GeV. We
also insure that the lightest chargino remains always heavier than
the lightest neutralino,  at least 15 GeV heavier, to avoid
significant contributions from co-annihilation channels. For the
mixing in the chargino sector, we assume all parameters (except
$M_R$  and $v_{\delta}$) large enough to obtain a pure right-wino
state. Using $v_\delta=30$ GeV, a 100 GeV or larger mass is
obtained for the right-wino,  while fulfilling the bound on the
mass of the lightest chargino. In order  to obtain a neutralino
lighter than the lightest chargino, $M_R$ should be at least $160$
GeV or larger; hence the excluded region (in green) in
Fig.~\ref{fig:omegaRWino}. For illustrative purposes only, 
we have included, in Table 1, the 
values of the masses of the LSP, as well as the masses of the 
second lightest neutralino, the lightest chargino, and the lightest sfermion,  for all three 
scenarios investigated. As in the case of pure bino, the
mass upper bound comes from sfermion sector. After
implementing WMAP constraint, the only surviving region at
$2\sigma$ level is shown in white strip.   This scenario satisfies
the WMAP conditions for a relatively large $M_{\rm SUSY}$ scale,
$600$ GeV or larger. Note that the upper bound for the mass of the
right wino is closely related to the chosen values for the other
gaugino masses, the vev's, and $\mu_2$, assumed heavy.  For
larger $M_R$ values, the lightest neutralino will not longer be a
pure state. But one can see from Fig.~\ref{fig:omegaRWino} that it
is not difficult to satisfy the WMAP result. Here too 
the constraints $b\to s\gamma$ and $(g-2)$ of the muon are taken into account
 with the parameter set,
$M_L=M_V=1.5$ TeV, $m_{\tilde{g}}=400$ GeV, $A=200$ GeV, $\mu_1=500$ GeV, 
$\mu_2=4$ TeV, $v_\delta=100$ GeV, and  $v_\Delta=1.5$ TeV. The $b\to s\gamma$
constraint for this case excludes mainly $M_{\rm SUSY}\le 600$ GeV region  
but practically doesn't effect the WMAP allowed region.  $(g-2)_\mu$ does not constrain the parameter space, as a light right-wino does not contribute significantly to the muon anomalous magnetic moment, and the other neutralinos are heavy. 

\begin{figure}[htb]
\vspace{-0.05in}
    \centerline{\epsfxsize 5.5in
{\epsfbox{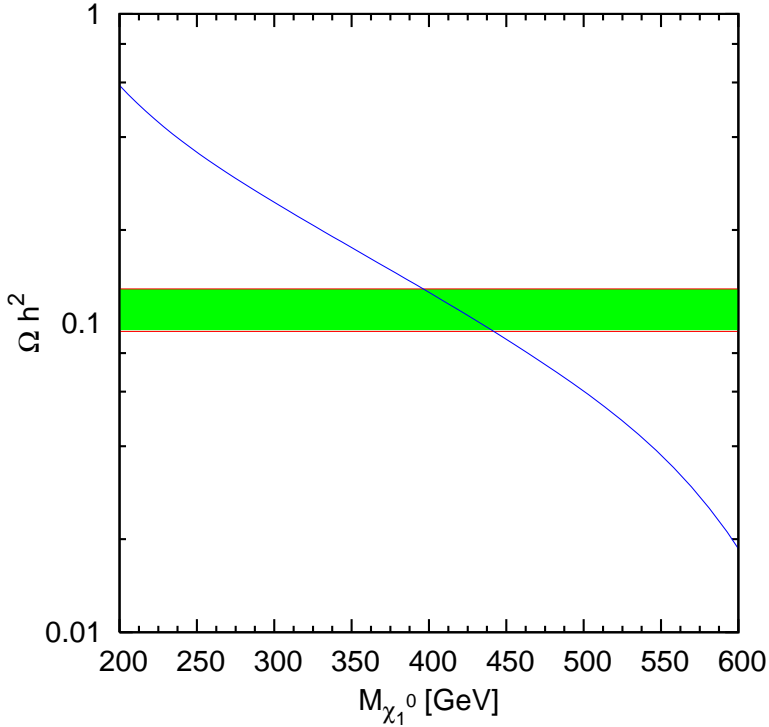}}}
\vskip -0.1in
 \caption{{\it The relic density of a pure higgsino $\tilde{\delta}_R^0$ as a function of its mass
for  $v_\delta=100$ GeV, $v_\Delta=1.5$ TeV, in the scenario in which the higgsino is the lightest neutralino.
The green strip is the allowed region by the WMAP at $2\sigma$ level.}}
 \label{fig:omegadelta}
\end{figure}
In the third scenario we investigate the possibility that the
$\tilde{\delta}_R^0$ is the lightest neutralino. This an interesting scenario, as $\tilde{\delta}_R^0$ is only introduced to cancel anomalies in the fermionic sector, and, because of its $B-L$ quantum number its direct interactions with matter are forbidden. The relic density prediction for such
scenario is shown in Fig.~\ref{fig:omegadelta}, as a function of
the  higgsino mass $M_{\tilde{\delta}_R^0}$, for $v_\delta=100$
GeV, $v_\Delta=1.5$ TeV. Only the $s$-channel contribution is 
available for the annihilation process
$\tilde{\delta}_R^0\bar{\tilde{\delta}_R^0}\stackrel{Z_R}{\to}
f\bar{f},\,\,(f=u,c,t,d,b,s...)$. Note that for this process, the
cross section and thus the relic density is formally independent
of $M_{\rm SUSY}$, since $\tilde{\delta}_R^0$ cannot interact with sfermions. The
$\Delta\rho$ bound doesn't allow values for $v_\Delta$ smaller
than $1.5$ TeV, once we assume a relatively small $v_\delta$. Then
the mass of the exchange particle $Z_R$ becomes heavy, around
$1.7$ TeV, yielding suppressed cross sections. This is why
we obtain a large relic density $\Omega_\chi
h^2|_{\chi=\delta_R^0}$,  which is inversely proportional to
thermally averaged cross section. Only larger masses for the pure higgsino lead to smaller
relic density values. In that case, as well as having more phase space available, one is closer to the $Z_R$ 
resonance, which increases the annihilation cross section. So, as seen from the figure, the WMAP at
$2\sigma$ level is satisfied if the mass of the higgsino lies in
400-450 GeV narrow range. Neither $b\to s \gamma$ nor
$(g-2)_\mu$ receive contributions from a $\tilde{\delta}_R^0$ higgsino. Unfortunately  the WMAP 
allowed region at $2\sigma$ level restricts the LSP $\tilde{\delta}_R^0$ mass to 
a very small interval. 

\begin{table}[htb]
	\caption{{\it Representative values for the masses of the LSP ($\chi_1^0$), the next-lightest neutralino ($\chi_2^0$), lightest chargino ($\chi_1^{\pm}$), and the lightest sfermion ($\tilde{f}_1$) for the three scenarios described in this section. The masses are given in GeV. The parameter sets are $(M_L,M_R,M_V,\mu_1,\mu_2,M_{\rm SUSY},v_\delta)=(1,1,0.13,1,2,0.2,0.05)$ TeV for Set (1), $(M_L,M_R,M_V,\mu_1,\mu_2,M_{\rm SUSY},v_\delta)=(1,0.2,1,1,4,0.8,0.01)$ TeV for Set (2), and $(M_L,M_R,M_V,\mu_1,\mu_2,M_{\rm SUSY},v_\delta)=(2,2,2,2,0.85,*,0.1)$ TeV for Set (3). For all three  sets, $\tan\beta=2$, $A=0.2$ TeV, and $v_\Delta=1.5$ TeV are used. "*" indicates that the entry depends on $M_{\rm SUSY}$ which can be taken arbitrarily large.}}
	\label{compare}
\begin{center}
   \begin{tabular*}{1\textwidth}{@{\extracolsep{\fill}} l l l l l l l l}
\hline
\hline
Scenario & LSP & $M_{\chi^0_1}$ &  $M_{\chi_2^0}$ & $M_{\chi_1^{\pm}}$ & $M_{\tilde{f}_1}$ & $\Omega_\chi h^2$ & Parameter Set\\
\hline
\hline
   1 &   bino                             &   130.6    &   472.2       &   500.0   &  176.2  &  0.106 & Set (1)\\
   2 &   R-wino                           &   177.1    &   924.5       &   196.4   &  713.4  &  0.121 & Set (2)\\
   3 &  $\tilde{ \delta}_R^0$ higgsino    &   412.5    &   1426.5      &   713.4   &   *     &  0.115 & Set (3)\\
\hline
\hline
\end{tabular*}
\end{center}
\end{table}

We will summarize our result in the next section.

\section{Conclusion}

In this study we have considered the neutralino sector of the LRSUSY model
and concentrated on the lightest neutralino state as dark matter,
motivated from the fact that WMAP result requires considerations of beyond 
MSSM models. Now, we would like to summarize our findings:

For the pure states in the LRSUSY model; if the
masses of the supersymmetric partners (in both bosonic and
fermionic sectors) are very small, the lightest neutralino is most
likely to be a pure $B-L$ bino. This covers regions with $M_{
\chi_1^0} \sim 60-220$ GeV, and $M_{\rm SUSY} \sim 150-250$ GeV.
For intermediate to larger values for the masses of the
supersymmetric partners, the lightest neutralino is most likely to
be pure right-wino; this covers regions with $M_{ \chi_1^0} \sim
150-350$ GeV, and $M_{\rm SUSY} \sim 600-1000$ GeV. Finally if the
mass of the lightest neutralino is large, in the $400-450$ GeV
region, the state could be pure $\tilde \delta_R^0$-higgsino. The
mass region is severely restricted for pure higgsino masses, but
completely independent of $M_{\rm SUSY}$, as long as $M_{\rm SUSY}
>500$ GeV, so that none of the sfermions are lighter than the
lightest neutralino.

The analysis above shows that the lightest neutralino as pure bino
shares some features with the similar scenario in MSSM. Both are
possible only if the scalars are light, especially for $M_V \sim
60$ GeV.
Outside this region, one obtains a relic density that is too
large. This similarity is not unexpected: although the bino here
is the fermionic partner of the $B-L$ gauge boson, and not the
hypercharge gauge boson, the same dominant decay channels are open
to both binos, so the cross sections are of the same order of
magnitude.

The situation changes when we analyze the case in which the
lightest neutralino is the right-wino. In MSSM, the case where the
lightest neutralino is the left-wino requires a left-wino mass of
$\sim 2.5$ TeV to satisfy dark matter limits. For our case,
right-wino masses in the $150-350$ GeV regions are in good
agreement with the relic density constraints. The reason for this
manifest difference is that the MSSM left-wino cross section is
dominated by the decay into $W_L^+W_L^-$ pairs, which is not
available for the right-wino.

The $s$-channel is available only for the $\tilde
\delta_R^0$-higgsino, but this channel is suppressed by the large
mass of the $Z_R$ boson propagator. The cross section is too small
for light neutralinos, and the relic density is too large. It is
less likely that the lightest neutralino would be the $\tilde
\delta_R^0$-higgsino, since only a very small mass range satisfies
the dark matter constraints. In MSSM,  the LSP is even less likely
to be the (MSSM) higgsino (see the recent study
\cite{Chattopadhyay:2005mv}). In LRSUSY, its annihilation cross
section into $W_L$ and $Z_L$ gauge boson pairs is unsuppressed,
and the relic density is too small, unless $\mu_1>1$ TeV.

Our scenarios are very different from those present in the NMSSM
or in other extensions of the MSSM, where a very light bino, or a
singlino, or their mixture could be the LSP
\cite{Belanger:2005kh,deCarlos:1997yv,Barger:2005hb}. There the
cross section is dominated by a light CP-odd Higgs boson or
additional Higgs resonances, and these particles only couple to no
SM particles except for the Higgs doublets.


\begin{acknowledgments}

The work of D. D. was partially supported by Turkish
Academy of Sciences through GEBIP grant, and by the Scientific and
Technical Research Council of Turkey through project 104T503. The work of M. F. and I. T. was funded in part  by NSERC of Canada (SAP0105354). The authors 
thank Goran Senjanovi{\v c} for useful e-mail exchange. M. F.
and I. T. would like to thank Genevi\`{e}ve B\'{e}langer for useful
comments. I. T. would also like to thank Marc Sher for helpful 
discussions. 
 \end{acknowledgments}

\appendix*
\label{sec:Appendixconstraints}

\section{Constraints }

We include, for completeness, the constraints on the parameter space coming from the anomalous magnetic moment of the muon $\Delta a_\mu$ and the branching ratio for $b \to s\gamma$. Constraints coming from $B_s \to \mu^+ \mu^-$ are more important than those coming from $b \to s \gamma$ only for $\tan \beta >11$ \cite{Frank:2002pm}, which does not affect the calculation for our choice of parameters. For larger $\tan \beta$ values, we expect the values for the masses of the supersymmetric partners to shift to larger values.

In addition to $b \to s\gamma$ and $\Delta a_\mu$, we have included constraints form the LEP limits on supersymmetric masses; we have constrained the mass of the next-to-lightest supersymmetric particle to be at least 15 GeV heavier than the lightest neutralino mass (to avoid co-annihilations); and we have restricted the Higgs vev's so the $\rho$ parameter:
\begin{eqnarray}
\rho^{-1}&=&\frac{M^2_{Z_L} \cos^2 \theta_W}{M^2_{W_L}}=1- \frac14 \frac{\cos^2 2 \theta_W}{\cos^4 \theta_W} \frac{\kappa_1^2 +\kappa_2^2}{v_{\Delta}^2+v_{\delta}^2}+{\cal O}\left[ \left (\frac{\kappa_1^2 +\kappa_2^2}{v_{\Delta}^2+v_{\delta}^2}\right)^2\right]
\end{eqnarray}
is consistent with its experimental limits $\rho=0.9998^{+0.0025}_{-0.0010}$ at 2$\sigma$ level \cite{Eidelman:2004wy}. 

\subsection{ $(g-2)_{\mu}$}
\label{subsec:amm}

In addition to the relic density, another experimental result that can constrain the $M_{1/2}-M_{\rm SUSY}$ parameter region is the BNL E821 measurement of the muon anomalous magnetic moment $a_{\mu} \equiv (g-2)_{\mu}/2$ \cite{Bennett:2004pv}:
\begin{equation}
a_{\mu}(\rm exp)=11~659~208(6) \times 10^{-10}
\end{equation}
which indicates that the anomalous magnetic moment may need additional contributions beyond the SM to be consistent with the experimental values. There are however uncertainties in how big the deviation from SM really is. A 2.4$\sigma$ deviation is obtained if the SM prediction of hadronic contributions from $e^+e^-$ data is used; if instead we use the data from indirect haronic $\tau$ decays the disagreement with the muon anomalous magnetic moment is reduced to a 0.9$\sigma$ deviation. We will use, from the 95\% C.L. range \cite{Lazarides:2006jw}
\begin{equation}
-0.53 \times 10^{-10} \le \Delta a_{\mu} \le 44.7 \times 10^{-10}
\end{equation}
The dominant contributions to $a_{\mu}$ in LRSUSY come from the chargino-muon sneutrino and neutralino-smuon loops. These are  \cite{Frank:1998ih}:
$a_{\mu}=a_{\mu}^{\chi^\pm}+ a_{\mu}^{\chi^0}$,   where:
\begin{eqnarray}
a_{L \mu}^{\chi^\pm} &=&\sum_{k=1}^4 \frac{m_\mu}{16 \pi^2}M_{\chi_k^{\pm}}g Y_{\mu}
Re[V_{k1}U_{k3}]
 \frac{F_3(x_{k\mu})}{m^2_{{\tilde \nu}_{\mu}}}
\end{eqnarray}
is the chargino contribution. The dominant neutralino contribution coming from  neutralino-left slepton graphs is:
\begin{eqnarray}
a_{L \mu}^{\chi^0} &=&\sum_{k=1}^7 \frac{m_\mu}{16  \pi^2}M_{\chi_k^{0}}g \left [\sqrt{2}  Y_{\mu}\left( N_{k1} + \tan^2 \theta_W
N_{k3} \right) N_{5 k} 
\frac {F_4(y_{k \mu_L}) }{m_{{\tilde \mu}_L}^2} \right. \nonumber \\
&+&\left. g  \left (N_{2 k}-2 \tan^2 \theta_W N_{k3}
\right) 
\left ( N_{k1} + \tan^2 \theta_W N_{k3} \right)  \right.
\nonumber \\
& \times & \left. \frac { m_{\mu} (A_{\mu}-\mu_1 \tan \beta) }{m_{{\tilde \mu}_R}^2} \frac {F_4(y_{k \mu_L}) }{m_{{\tilde \mu}_L}^2}\right]
\end{eqnarray}
and from the neutralino-right slepton graphs:
\begin{eqnarray}
a_{R \mu}^{\chi^0} & = &\sum_{k=1}^7\frac{m_{\mu}}{16  \pi^2} M_{\chi_k^{0}}g \left[ \frac{1}{\sqrt{2} }  Y_{\mu} \left( N_{k2} -2
\tan^2
\theta_W N_{k3} \right) N_ {k5} \frac {F_4(y_{k \mu_R}) }{m_{{\tilde \mu}_R}^2} \right. \nonumber \\
& + &\left. g\left( N_{k1} + \tan^2 \theta_W
N_{k3} \right) \left (N_{k2}-2 \tan^2 \theta_W N_{k3} \right) \right. \nonumber \\
& \times  &\left.   \frac{ m_{\mu} (A_{\mu}-\mu_1 \tan \beta) }{m_{{\tilde \mu}_L}^2} \frac {F_4(y_{k \mu_R}) }{m_{{\tilde \mu}_R}^2} \right]
\end{eqnarray}
with the loop functions $F_3(x),\, F_4(x)$ are given in Eq. (\ref{loopfunctions}) and where $x_{k\mu}=\displaystyle{\frac {M_{\chi_k^\pm}^2}{m_{\tilde {\nu}_{\mu}}^2}}$,  $y_{k\mu_{L,R}}=\displaystyle{\frac {M_{\chi_k^0}^2}{m_{\tilde {\mu}_{L,R}}^2}}$, and $U,\, V$ (and $N$) are matrices that diagonalize the chargino (and neutralino) mass matrices.

\subsection{$b\to s \gamma$}
\label{subsec:bsg}

The inclusive decay width for the process $b \rightarrow s \gamma$ is
given by
\begin{equation}
\Gamma(b \rightarrow s \gamma)=\frac{m_b^5 G_F^2 
\alpha}{32 \pi^4}
\left( \hat{M}_{\gamma L}^2+\hat{M}_{\gamma R}^2
\right),
\end{equation}
where $\hat{M}$ means evolving down to the decay scale $\mu=m_b$.
The branching ratio can be expressed as
\begin{equation}
BR (b\rightarrow s \gamma)= \frac{\Gamma (b\rightarrow s
\gamma)}{\Gamma_{SL}} BR_{SL},
\end{equation}
where the semileptonic branching ratio $BR_{SL}=BR(b \rightarrow ce
{\bar \nu})=(10.49
\pm 0.46)\%$ and
\begin{equation}
\Gamma_{SL}=\frac{m_b^5 G_F^2 |(K_{CKM})_{cb} |^2 }{192 \pi^3}g(z),
\end{equation}
where $z=m_c^2/m_b^2$ and $g(z)=1-8z+8z^3-z^4-12z^2 \mathrm{log}z$. The experimental measurement from CLEO can be expressed as \cite{Chen:2001fj}:
\begin{equation}
BR( b \to s \gamma)=(3.21 \pm 0.43 \pm 0.27^{+ 0.18}_{-0.10}) \times 10^{-4}
\end{equation}
The SM contribution is 
\begin{eqnarray}
A_{SM}=\frac{\pi\alpha_W}{2 \sqrt{2} G_F M_W^2} (K_{CKM})_{ts}^{*} (K_{CKM})_{tb}\,3x_{tW}\left [Q_uF_1(x_{tW})+F_2(x_{tW})\right].
\end{eqnarray}

The matrix elements responsible for the $b \rightarrow s \gamma$
decay acquire the
following contributions from the supersymmetric sector of the model \cite{Frank:2002nj}.
For $b_L$ decay:
\begin{equation}
M_{\gamma_R}=A_{\tilde{g}}^R+A_{\chi^-}^R+A_{\chi^0}^R
\end{equation}
with the gluino, chargino and neutralino contributions given by
\begin{eqnarray}
A_{\tilde{g}}^R &=& - \frac{\sqrt{2}\pi \alpha_s}{ G_F} Q_d C(R)
\sum_{k=1}^6
\frac{1}{m_{\tilde{d}_k}^2} \{ \Gamma_{DL}^{kb} \Gamma_{DL}^{*ks}
F_2(x_{\tilde{g} \tilde{d_k}})
-\frac{m_{\tilde{g}}}{m_b} \Gamma_{DR}^{kb} \Gamma_{DL}^{*ks}
F_4(x_{\tilde{g} \tilde{d_k}}) \}, \\
A_{\chi^-}^R &=& - \frac{\pi \alpha_W}{\sqrt{2} G_F}
\sum_{j=1}^5 \sum_{k=1}^6
\frac{1}{m_{\tilde{u}_k}^2} \{
(G_{UL}^{jkb}-H_{UR}^{jkb})(G_{UL}^{*jks}-H_{UR}^{*jks})
[ F_1(x_{\chi_j \tilde{u}_k})+ Q_u F_2(x_{\chi_j
\tilde{u}_k})] \nonumber \\
    & & +\frac{m_{\chi_j}}{m_b} (G_{UR}^{jkb}-H_{UL}^{jkb})
(G_{UL}^{*jks}
-H_{UR}^{*jks}) [F_3 (x_{\chi_j \tilde{u}_k})+ Q_u
F_4(x_{\chi_j \tilde{u}_k})] \}, \\
A_{\chi^0}^R &=& - \frac{\pi \alpha_W}{\sqrt{2} G_F} Q_d
\sum_{j=1}^9 \sum_{k=1}^6
\frac{1}{m_{\tilde{d}_k}^2} \{
(\sqrt{2}G_{0DL}^{jkb}-H_{0DR}^{jkb})(\sqrt{2}G_{0DL}^{*jks}-H_{0DR}^{*jks})
F_2(x_{\chi_j^0 \tilde{d}_k}) \nonumber \\
    & & +\frac{m_{\chi_j^0}}{m_b}
(\sqrt{2}G_{0DR}^{jkb}-H_{0DL}^{jkb}) (\sqrt{2} G_{0DL}^{*jks}
-H_{0DR}^{*jks})  F_4(x_{\chi_j^0 \tilde{d}_k}) \},
\end{eqnarray} 
and for the decay of $b_R$:
\begin{equation}
M_{\gamma_L}=A_{\tilde{g}}^L+A_{\chi^-}^L+A_{\chi^0}^L
\end{equation}
again, with the following gluino, chargino and neutralino contributions
\begin{eqnarray}
A_{\tilde{g}}^L &=& - \frac{\sqrt{2}\pi \alpha_s}{ G_F} Q_d C(R)
\sum_{k=1}^6
\frac{1}{m_{\tilde{d}_k}^2} \{ \Gamma_{DR}^{kb} \Gamma_{DR}^{*ks}
F_2(x_{\tilde{g} \tilde{d_k}})
-\frac{m_{\tilde{g}}}{m_b} \Gamma_{DL}^{kb} \Gamma_{DR}^{*ks}
F_4(x_{\tilde{g} \tilde{d_k}}) \}, \\
A_{\chi^-}^L &=& - \frac{\pi \alpha_W}{\sqrt{2} G_F}
\sum_{j=1}^5 \sum_{k=1}^6
\frac{1}{m_{\tilde{u}_k}^2} \{
(G_{UR}^{jkb}-H_{UL}^{jkb})(G_{UR}^{*jks}-H_{UL}^{*jks})
[ F_1(x_{\chi_j \tilde{u}_k})+ Q_u F_2(x_{\chi_j
\tilde{u}_k})] \nonumber \\
    & & +\frac{m_{\chi_j}}{m_b} (G_{UL}^{jkb}-H_{UR}^{jkb})
(G_{UR}^{*jks}
-H_{UL}^{*jks}) [F_3 (x_{\chi_j \tilde{u}_k})+ Q_u
F_4(x_{\chi_j \tilde{u}_k})] \}, \\
A_{\chi^0}^L &=& - \frac{\pi \alpha_W}{\sqrt{2} G_F} Q_d
\sum_{j=1}^9 \sum_{k=1}^6
\frac{1}{m_{\tilde{d}_k}^2} \{
(\sqrt{2}G_{0DR}^{jkb}-H_{0DL}^{jkb})(\sqrt{2}G_{0DR}^{*jks}-
H_{0DL}^{*jks})
F_2(x_{\chi_j^0 \tilde{d}_k}) \nonumber \\
    & & +\frac{m_{\chi_j^0}}{m_b}
(\sqrt{2}G_{0DL}^{jkb}-H_{0DR}^{jkb}) (\sqrt{2} G_{0DR}^{*jks}
-H_{0DL}^{*jks})  F_4(x_{\chi_j^0 \tilde{d}_k}) \},
\end{eqnarray}
where the
chargino-quark-squark mixing martices $G$
and $H$ are defined as
\begin{eqnarray}
G^{jki}_{UL} &=& V_{j1}^{\ast}  (\Gamma_{UL})_{ki},
\nonumber \\
G^{jki}_{UR} &=& U_{j2}  (\Gamma_{UR})_{ki}, \nonumber \\
H_{UL}^{jki} &=& \frac{1}{\sqrt{2} m_W} ( \frac{m_{u_l}}{\sin \beta}(K_{CKM})_{il}
U_{j3}+
\frac{m_{d_l}}{\cos \beta} \delta_{il}U_{j4} )  (\Gamma_{UL})_{kl},
\nonumber \\
H_{UR}^{jki} &=& \frac{1}{\sqrt{2} m_W} ( \frac{m_{u_l}}{\sin \beta}(K_{CKM})_{il}
V_{j3}^*+
\frac{m_{d_l}}{\cos \beta} \delta_{il}V_{j4}^* ) (\Gamma_{UR})_{kl}.
\end{eqnarray}
and where  the neutralino-quark-squark mixing matrices $G_0$
and $H_0$ are defined as
\begin{eqnarray}
G^{jki}_{0DL} &=&\left \{  - \left [ \tan^2 \theta_W \left (T_d^3-Q_d-\frac {Q_u+Q_d}{2}\right)
+\frac {Q_u+Q_d}{2}\right]  N^{\ast}_{j1}\right. \nonumber \\
& -& \left. \frac{\sqrt{\cos 2 \theta_W }}{\cos \theta_W} \tan \theta_W \left (T_d^3-Q_d-\frac {Q_u+Q_d}{2}\right)
 N^{\ast}_{j2}+ T_d^3N^{\ast}_{j3} \right \} (\Gamma_{DL})_{ki}, \nonumber \\
G^{jki}_{0DR} &=& \left\{  - \frac {\cos 2\theta_W}{\cos^2 \theta_W}\left (  T^3_d-2Q_d \sin^2
\theta_W \right )N_{j1}  \right. \nonumber \\
&+&\left. \sqrt{\cos 2\theta_W}\sin \theta_W \left [ Q_d+\frac{1}{\cos^2\theta_W} \left (T^3_d-2 Q_d \sin^2 \theta_W\right) \right] N_{j2}+ 2\sin^2 \theta_W Q_d N_{j3}\right \}
(\Gamma_{DR})_{ki},  \nonumber \\
H_{0DL}^{jki} &=&  \frac{1}{\sqrt{2} m_W} ( \frac{m_{u_l}}{\sin \beta}
N_{j4}-
\frac{m_{d_l}}{\cos \beta} N_{j5}) 
(\Gamma_{DL})_{ki}, \nonumber \\
H_{0DR}^{jki} &=& \frac{1}{\sqrt{2} m_W} ( \frac{m_{u_l}}{\sin \beta}
N^{ \ast}_{j4}+
\frac{m_{d_l}}{\cos \beta} N^{ \ast}_{j5} ) (\Gamma_{DR})_{ki}.
\end{eqnarray}
and where the matrices $\Gamma_{U,D}$ diagonalize the squark mass matrices in the up and down  sectors.
The functions appearing in the expressions above are 
\begin{eqnarray}
\label{loopfunctions}
F_1(x)&=&\frac{1}{12(x-1)^4}(x^3-6x^2+3x+2+6x\log x),\nonumber \\
F_2(x)&=&\frac{1}{12(x-1)^4}(2 x^3+3x^2-6x+1-6x^2\log x), \nonumber \\
F_3(x)&=&\frac{1}{2(x-1)^3}(x^2-4x+3+2\log x),\nonumber \\
F_4(x)&=&\frac{1}{2(x-1)^3}(x^2-1-2x\log x)
\end{eqnarray}
The convention
$x_{ab}=m_a^2/m_b^2$ is used. $C(R) =4/3$ is the quadratic Casimir
operator of the fundamental
representation of $SU(3)_C$.

In order to compare the results obtained with experimental branching
ratios, QCD
corrections must be taken into account.
We assume the renormalization group evolution pattern.
 There is no mixing between left and
right-handed contributions.
\begin{equation}
A^{\gamma} (m_b)=\eta^{-16/23} \{ A^{\gamma} (M_W) + A^{\gamma}_0 [
\frac{116}{135} (\eta^{10/23}-1)
+\frac{58}{189}(\eta^{28/23}-1)] \},
\end{equation}
where $\eta=\displaystyle{ \alpha_s(m_b)/ \alpha_s(M_W)}$ and $A^{\gamma}_0= \displaystyle{\frac{\pi
\alpha_W } {2 \sqrt{2} G_F M_W^2} (K_{CKM})_{ts}^{*}(K_{CKM})_{tb}}$. We choose the renormalization scale to be $\mu=m_b=4.2$ GeV.


\end{document}